\documentclass[a4paper,11pt]{article}
\pdfoutput=1 % if your are submitting a pdflatex (i.e. if you have
\newlength{\tmplengtha}
\newlength{\tmplengthb}
\newlength{\tmplengthc}

\newcommand{\eqspace}{\hphantom{={}}\:}

\usepackage{environ}
\NewEnviron{boxalign*}{
   \begin{center}
   \begin{minipage}{\textwidth}
   \fbox{
     \begin{minipage}{\textwidth-4\fboxsep}
     \vspace{-3.5\fboxsep}
     \begin{align*}
       \BODY
     \end{align*}
     \end{minipage}
   }
   \end{minipage}
   \end{center}
}

\NewEnviron{boxalign}[1]{
     \begin{center}
           \begin{minipage}{\textwidth}
               \fbox{
                 \begin{minipage}{\textwidth-4\fboxsep}
                     \vspace{-3.5\fboxsep}
                     \begin{align*}
                           \BODY
                     \end{align*}
                 \end{minipage}
               }
               \vspace{-4\fboxsep}
               \begin{align}
                   \label{#1}
               \end{align}
           \end{minipage}
       \end{center}
}
\usepackage[a4paper,textwidth=460.72124pt,textheight=621.0pt]{geometry}
\usepackage{cite}
\usepackage{graphicx}
\usepackage[labelfont=bf]{caption}
\usepackage{subcaption}
\usepackage[T1]{fontenc} 
\usepackage[utf8]{inputenc}

\usepackage{amssymb}
\usepackage{amsmath}
\usepackage{amsfonts}
\usepackage{mathtools}
\usepackage{etoolbox} % for simple if constructs
\usepackage{booktabs} % nicer tables

\DeclareMathAlphabet{\mathpzc}{OT1}{pzc}{m}{it}
\numberwithin{equation}{section}

\usepackage[colorlinks]{hyperref}
\hypersetup{
 citecolor=blue,
 linkcolor=blue,
 urlcolor=blue}

\usepackage[capitalize]{cleveref}
\usepackage{physics}
\usepackage{color,soul}
\newcommand{\ep}{\epsilon}
\newcommand{\be}{\begin{equation}}
\newcommand{\ee}{\end{equation}}
\newcommand{\bea}{\begin{eqnarray}}
\newcommand{\eea}{\end{eqnarray}}
\begin{document}

%>>> front page
\pagenumbering{roman}
\pagestyle{empty}
\vspace{-5.0cm}
\begin{flushright}
  P3H-20-006, TTP20-004
\end{flushright}

\vspace{2.0cm}
\begin{center}
  {\large \textbf{
    NNLO zero-jettiness beam and soft functions to higher orders in the dimensional-regularization parameter $\epsilon$}}\\
\end{center}

\vspace{0.5cm}
\begin{center}
  Daniel Baranowski$^{a}$,
  \\
  \vspace{0.3cm}
  {\textit{
    {}$^{a}$ Institut f{\"u}r Theoretische Teilchenphysik (TTP), KIT, 76128 Karlsruhe, Germany\\
  }}
\end{center}

\vspace{1.3cm}
\begin{center}
{\large \textbf{Abstract} }
\end{center} {
  We present the calculation of the next-to-next-to-leading order (NNLO)
  zero-jettiness  beam and soft functions, up to the
  second order in the expansion in the dimensional regularization parameter $\epsilon$. These higher order 
  terms are needed for the computation of the
  next-to-next-to-next-to-leading order (N$^3$LO) zero-jettiness soft and beam functions. As a byproduct, we confirm the $\order{\epsilon^0}$ results for NNLO beam and soft functions available in the literature~\cite{Tackmann2014,Gaunt:2014cfa,Petriello,Monni:2011gb,Kelley:2011ng}.
}
\clearpage

%>>> table of contents
{
  \hypersetup{linkcolor=black}
  \tableofcontents
}
\clearpage

%>>> text
\pagestyle{plain}
\pagenumbering{arabic}
\allowdisplaybreaks

%-------------------------
\section{Introduction}
To find signals of physics beyond the Standard Model, many interesting 
processes at the LHC are being studied  with ever increasing precision.
An important part of these efforts is the development of methods that enable N$^3$LO QCD calculations,
at least for the simplest processes where color-singlet final states
are produced. In the absence of fully-developed N$^3$LO subtractions schemes, 
a promising approach is the slicing method~\cite{Catani:2007vq,N2LO4,Gaunt:2015pea,Bonciani:2015sha} that has seen a recent
resurgence in the context of LHC physics.

Any slicing method is based on the idea that one can split the phase space for a process of interest into partially-resolved and
fully-unresolved parts. The fully-unresolved contribution originates from virtual, real-soft and real-collinear emissions. Conversely, the
resolved one requires a final state that contains at least one additional QCD jet in comparison to the lowest order final state and, for this reason, 
it must be computed through  lower order in the perturbative expansion in  QCD than the unresolved one. 

Phase-space separation into fully-unresolved and resolved parts can be accomplished using different kinematic variables.  The two most popular
ones are $p_\perp$ and $N$-jettiness  variables that have been used recently in many NNLO QCD computations~\cite{Catani:2009sm,N2LO1,Boughezal:2015ded,Boughezal:2016wmq,Boughezal:2016dtm,Grazzini:2016ctr,Grazzini:2016swo,Grazzini:2017mhc,Grazzini:2017ckn,Catani:2018krb,Catani:2019hip,Boughezal:2019ggi}. In this paper
we will deal with the so-called zero-jettiness variable that can be used to perform a slicing computation
of N$^3$LO QCD corrections to the production of a colorless final state $V$~($H$, $W$, $Z$, $\gamma^*$, $WW$, $ZZ$, $\gamma \gamma$, etc.) in hadron collisions.  This variable reads \cite{Stewart:2010tn} 
 \begin{align}
    \tau=\sum_{m} \min_{ \mathrm{i} \in \{1,2\}} \left[\frac{2 p_i \cdot k_m}{Q_i} \right] \label{eq:sliceagain},
\end{align}
where $p_i$ are the four-momenta of incoming partons, $k_m$ are the momenta of final state QCD partons and 
$Q_i$ are the so-called hardness variables. In the limit of small $\tau$, the cross section factorizes~\cite{Stewart:2009yx}
into a product of
hard $H$, beam $B$ and soft $S$ functions
\begin{align}
  \lim_{\mathcal \tau_0 \to 0}  \dd[]\sigma^{\mathrm{N^3LO}}_{pp\to V+X} \ \left(\tau \textless \mathcal \tau_0 \right)=
  B \otimes B \otimes S \otimes H \otimes \dd[] \sigma^{\mathrm{LO}}_{pp\to V} \ . \label{eq:split2}
\end{align}

All quantities that appear in \cref{eq:split2} are known through NNLO QCD. Moreover, the hard function $H$
is known through N$^3$LO QCD for single vector boson and Higgs boson production~\cite{Baikov:2009bg,Gehrmann:2010ue}
and, recently, the three-loop quark-to-quark  matching coefficient, needed to relate the beam function
to parton distribution functions, was computed in the generalized large-$N_c$ approximation in Ref.~\cite{Behring:2019quf}. \footnote{ We note that
the computations of the N$^3$LO QCD quark-to-quark, gluon-to-quark and anti-quark-to-quark matching coefficients for $p_\perp$ variable were reported in Ref.~\cite{Luo:2019szz}.}
The computation reported in Ref.~\cite{Behring:2019quf} required the knowledge of certain 
NNLO beam functions through second order in the dimensional regularization parameter $\epsilon$. These functions were calculated in Ref.~\cite{Baranowski} and the results of that
computation were used in Ref.~\cite{Behring:2019quf}.  

The goal of this paper is twofold.
First, we aim to extend the calculation reported in Ref.~\cite{Baranowski} 
and to compute {\it all} NNLO QCD matching coefficients
 through the second order in $\epsilon$, as required for the calculation of matching coefficients through N$^3$LO QCD. Second, we will compute
  the NNLO QCD soft function through the second order in $\epsilon$, as required for the calculation of the N$^3$LO QCD soft function.
We note that NNLO QCD zero-jettiness beam functions were computed in 
Refs.~\cite{Tackmann2014,Gaunt:2014cfa,Petriello} through zeroth order in $\epsilon$, whereas the NNLO soft function was originally calculated in Refs.~\cite{Monni:2011gb,Kelley:2011ng}.

To extend the calculation of beam and soft functions to higher orders in $\epsilon$, we use methods that may be of interest in their own right.
Indeed,  we employ
collinear and soft limits of QCD amplitudes~\cite{Catani:1999ss,Catani:2000pi}, reverse unitarity~\cite{reverseunit}
and integration-by-parts identities~\cite{ibp} to show  that
computation of soft and {\it all} NNLO beam functions for zero-jettiness can be significantly simplified. In the case of the soft function, we rewrite step functions that arise from the definition of the zero-jettiness variable
as integrals of delta functions over auxillary parameters before applying reverse unitarity. We note that these methods allow one to express any NNLO zero-jettiness beam function through just 
{\it twelve} and the NNLO soft function through just {\it nine} simple (phase-space or loop) integrals.
In case of the soft function, integrations over auxillary parameters turn out to be remarkably simple.

The remainder of the  paper is organized  as follows. 
In \cref{sec:calcset} we  describe the computation of  the partonic beam functions through ${\cal O}(\epsilon^2)$  starting
from collinear limits of scattering amplitudes and explain how the master integrals are calculated. We discuss the calculation of the bare soft function through ${\cal O}(\epsilon^2)$ in
 \cref{sec:soft}. 
We conclude in \cref{sec:con}.
Finally, we note that results for the NNLO bare soft function and beam function matching coefficients are collected in an ancillary file provided with this submission.

%-------------------------

%-------------------------
\section{Calculation of the beam function}
\label{sec:calcset}
In this section we describe the calculation of the bare partonic beam function. We
split the discussion into two parts. In \cref{subsec:bbf}
we explain the general set up and relate the calculation of the beam functions to collinear limits of QCD amplitudes. We also use reverse
unitarity to express bare beam functions through master integrals.
In \cref{subsec:micalc} we describe the calculation of these master integrals. We present some results
in \cref{subsec:resultsbeam}.
\subsection{General setup}
\label{subsec:bbf}
It was pointed out in Ref.~\cite{Waalewijn} that a bare partonic beam function $B^{b}_{ij}$, that describes the transition of a parton $j$ to a parton $i$, can be
obtained by integrating spin- and color-averaged collinear splitting functions $ \left\langle  P_{j\to i^* \{m\}} \right\rangle$ over an unresolved $m$-particle phase-space
\begin{align}
    B^b_{ij}\sim \sum_{\{m\}} \int \dd[] \mathrm{PS^{(m)}} \left\langle  P_{j\to i^* \{m\}} \right\rangle.\label{eq:splitting}
\end{align}
The phase-space measure is defined as follows
\begin{align}
    \dd[]\mathrm{PS^{(m)}}=\left(\prod_n^m \frac{\dd[d]{k_n}}{{(2 \pi)}^{d-1}}  \  \delta^+ \left( k_n^2 \right) \right) \delta\left( 2 \sum_n^m k_{n} \cdot p - \frac{t}{z} \right)   \delta \left( 2\sum_n^m\frac{ k_{n} \cdot \bar{p}}{s} - (1-z)\right) , \label{eq:PSm}
\end{align}
where $\{m\}$ is the set of collinearly-radiated partons.
In \cref{eq:PSm}
we denote the momentum of the incoming parton $j$ as $p$, its complementary light-cone momentum as $\bar{p}$ and the momenta of final state partons as $k_m$.
Furthermore,
$t$ is the so-called transverse virtuality of the off-shell parton $i$, $z \cdot p$ is its longitudinal momentum and $s=2 p\cdot\bar{p}$.
It was explained in Ref.~\cite{Catani:1999ss} how
splitting functions $P_{j \to i^*}$ for all parton-to-parton transitions can be calculated. This requires the use of a physical (axial) gauge
for gluons and projection operators that decouple collinear emissions from hard matrix elements.
These projection operators act on matrix elements $M_{j\to i^* \{m\}}$ describing the process of a parton $j$ splitting into on-shell partons $\{m\}$
 and an off-shell parton $i^*$.

\begin{figure}[!b]
    \centering
    \begin{subfigure}{.45\textwidth}
      \centering
      \includegraphics[width=\linewidth]{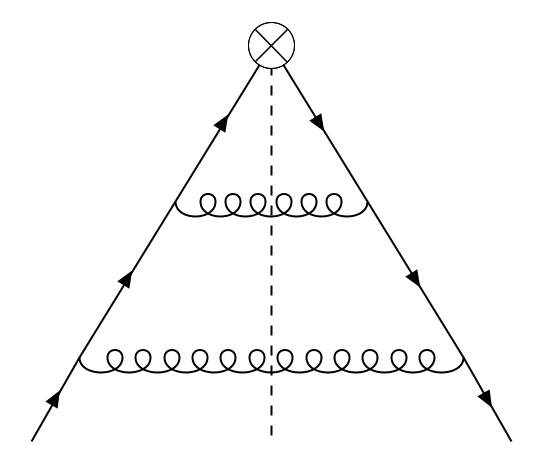}
      \caption{\;}
    \end{subfigure}%
    \hfil
    \begin{subfigure}{.45\textwidth}
      \centering
      \includegraphics[width=\linewidth]{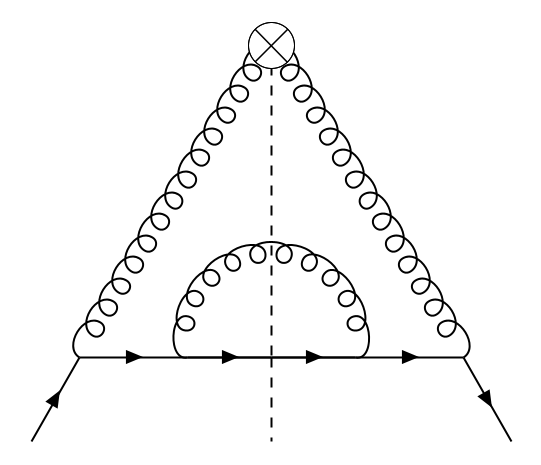}
      \caption{\;}
    \end{subfigure}
    \hfil
    \caption{Example diagrams contributing to the $B_{qq}$(a) and $B_{gq}$(b) beam functions. The dashed line represents a ``cut'' so that all particles crossing it are on the mass-shell.
    The vertex $\otimes$ denotes the insertion of the projection operator defined in \cref{eq:proj1}.}
  \label{fig:exp}
    \end{figure}

Following Ref.~\cite{Catani:1999ss}, we write
\begin{align}
  \left\langle  P_{j\to i^* \{m\}} \right\rangle&=\mathcal{P}|M_{j\to i^* \{m\}}|^2,\label{eq:proj} \\
    \mathcal{P}|M_{j\to i^*\{m\}}|^2&= \left\{
    \begin{array}{ll}
        \displaystyle
         \sum \;	\Tr[M_{j\to i^* \{m\}}\frac{\hat{\bar{p}}}{4 \bar{p} \cdot p_s}M^{\dagger}_{j\to i^* \{m\}}],& \mathrm{if }\ i \in \{q,\bar{q}\}  \label{eq:proj1}\\
         \displaystyle -\frac{1}{2(1-\epsilon)} \sum \;  d^\rho_\mu\left(p_s \right)d_{\nu\rho}\left(p_s \right)  M^{\mu}_{j\to i^* \{m\} } M^{\nu \dagger }_{j\to i^* \{m\}} ,& \mathrm{if }\ i \in\{g\}
    \end{array} \right.
    \end{align}
where 
\begin{align}
   d_{\mu \nu}(k) &=-g_{\mu \nu}+\frac{k_\mu \bar{p}_\nu + \bar{p}_\mu k_\nu }{ \ k \cdot \bar{p}}, &p_s&=p-\sum_m k_m,
\end{align}
 and the sums in \cref{eq:proj1} run over color, polarization and spin degrees of freedom of all external particles.
Combining \cref{eq:splitting} and \cref{eq:proj}, we write the beam function as
\begin{align}
    B^b_{ij}=\sum_{\{m\}} \frac{1}{\mathcal{N}_m} \int \dd[] \mathrm{PS^{(m)}} \mathcal{P}|M_{j\to i^* \{m\}}|^2,\label{eq:Matrixsum}
\end{align}
where $\mathcal{N}_m$ are symmetry and averaging factors.
To compute all beam functions it is sufficient to consider $i$'s and $j$'s from the following set $\left(i,j \right)\in \{(q_l,q_m),(q_l,g),(q_l,\bar{q}_m),(g,g),(g,q_m)\}$\cite{Tackmann2014,Gaunt:2014cfa}, where the indices $l$ and $m$ denote quark flavours.
We note that a flavour-preserving transition in $B^b_{q_lq_m}$ is obtained by setting $l=m$.
Similar to regular splitting functions, all other beam functions can be obtained from the above set.
Examples of diagrams  that are required for the calculation of beam functions are shown\footnote{We use FeynGame~\cite{Harlander:2020cyh} to draw Feynman diagrams.} in \cref{fig:exp}.

% S
The bare partonic beam functions $B^b_{ij}$ \cref{eq:Matrixsum} can now be calculated as standard phase-space and loop integrals with the projection operator $\mathcal{P}$ as a special Feynman rule.
To facilitate this computation, we apply reverse unitarity\cite{reverseunit} and rewrite delta functions in \cref{eq:PSm} as differences of two ``propagators'' with opposite signs in the $i0$ prescription,
mapping phase-space integrals in \cref{eq:Matrixsum} onto loop integrals. We then use integration-by-parts (IBP) identities \cite{ibp} to express the beam function through master integrals. The IBP reduction
  is performed using FIRE \cite{fire}. 

We find that all five beam functions can be expressed through just 12 master integrals.
They include nine double-real master integrals
\begin{align}
  &I_1=  \left[  1 \right]_{(2)} ,    
  &I_2&=  \left[  \frac{ 1 }{\bar{p}\cdot (p-k_1)} \right]_{(2)},  \notag  \\
   &I_3= \left[  \frac{1}{(p-k_{12})^2}\right]_{(2)},
   &I_4&= \left[  \frac{ 1 } {(p-k_1)^2  \ k_{12}^2  \ \bar{p}\cdot k_2}\right]_{(2)} , \notag \\
   & I_5= \left[  \frac{1}{(p-k_1)^2 \ (p-k_{12})^2 \ \bar{p}\cdot k_1} \right]_{(2)},
    &I_6&= \left[  \frac{1} {(p-k_1)^2  \ (p-k_{12})^2 \ \bar{p}\cdot k_2}\right]_{(2)},  \label{eq:RRmasters} \\  
   &I_7= \left[  \frac{1}{(p-k_{12})^2 \  [\bar{p}\cdot (p-k_1)]}\right]_{(2)},  &I_8&= \left[   \frac{1}{k_{12}^2 \ [ \bar{p}\cdot (p-k_1)] \ (p-k_1)^2}\right]_{(2)},  \notag \\
   &I_9= \left[  \frac{1}{(p-k_{12})^2 \ (p-k_2)^2 \ (p-k_1)\cdot\bar{p}}\right]_{(2)}, &&  \notag
\end{align}
  and three  real-virtual master integrals
  \begin{align} 
    I_{10}=& \left[ \frac{1}{l^2 \ l \cdot \bar{p}\ (p-l)^2 \ (p-k-l)^2}\right]_{(1)}, 
    &I_{11}=&\left[  \frac{1}{l^2 \ (p-k-l)^2}\right]_{(1)} ,\label{eq:RVmasters} \\
   I_{12}=&\left[  \frac{1}{l^2 \ (p-l)^2 \ (l-k)^2 \ (l-k) \cdot \bar{p}}\right]_{(1)},
   && \notag
  \end{align}
where for a given integrand $f$ we write
  \begin{align}
    &\left[  f \right]_{(2)}=\int\dd[]\mathrm{PS^{(2)}} \; f, &\left[  f \right]_{(1)}=\int\dd[]\mathrm{PS^{(1)}} \int \frac{\dd[d]l}{(2 \pi)^{d}} \; f.
  \end{align}
  We note that phase-space measures $\dd[]\mathrm{PS^{(2,1)}}$ are defined in \cref{eq:PSm}.
  We also note that the gluon and quark beam function share the same set of master integrals.
  We describe the calculation of these master  integrals in the next section.

  \subsection{Master integrals}
\label{subsec:micalc}
The master integrals shown in \cref{eq:RRmasters,eq:RVmasters} are sufficiently simple to be evaluated directly.
To illustrate the computation, we discuss three representative examples. All other master integrals can be calculated along similar lines.

%1
  We begin with the master integral
  \begin{align}
    I_6&= \int \frac{\dd[d]k_1}{(2 \pi)^{d-1}} \int \frac{\dd[d]k_2}{(2 \pi)^{d-1}} \ \delta^+\left( k_1^2\right ) \delta^+\left( k_2^2\right )  \delta\left( 2 k_{12} \cdot p -  \tfrac{t}{z} \right) \frac{ \delta \left( \tfrac{ 2 k_{12} \cdot \bar{p}}{s} - (1-z) \right) } {(p-k_1)^2  \ k_{12}^2  \ \bar{p}\cdot k_2}.
  \end{align}
 We start by rescaling the momenta $p$, $\bar{p}$, $k_1$ and $k_2$ in such a way that the dependencies of the integrals on $t$ and $s$ factor out.
To this end, we write\footnote
    {We note that for real-virtual master integrals we also rescale the loop momentum
    $
      l\to\tilde{l}\sqrt{t}.
    $}
 \begin{align}
   &\bar{p}=\tilde{\bar{p}}\frac{s}{\sqrt{t}}, 
   &&p=\tilde{p} \sqrt{t},  
   &&k_i=\tilde{k}_i \sqrt{t},  
 \end{align}
and obtain
 \begin{align}
     I_6 \left(s,t,z \right)=t^{d-5}s^{-1}I_6 \left(1,1,z\right).
 \end{align}
 To simplify the notation we drop tildes over momenta and turn to the calculation of the following integral
 \begin{align}
    I_6 \left(1,1,z\right)&= \int \frac{\dd[d]k_1}{(2 \pi)^{d-1}} \int \frac{\dd[d]k_2}{(2 \pi)^{d-1}} \ \delta^+\left( k_1^2\right ) \delta^+\left( k_2^2\right )\delta\left( 2 k_{12} \cdot p -  \tfrac{1}{z} \right) \frac{ \delta \left( 2 k_{12} \cdot \bar{p} - (1-z) \right) } {(p-k_1)^2  \ k_{12}^2  \ \bar{p}\cdot k_2} . \label{eq:Dmaster}
\end{align}
We insert \(1=\int \dd[d]Q \ \delta^d(k_1+k_2-Q) \) into the integrand and change the order of integration. We find
 \begin{align}
    I_6 \left(1,1,z\right)&= \int \dd[d]Q \ \delta\left( 2 Q \cdot p - \tfrac{1}{z} \right) \delta \left( 2 Q \cdot \bar{p} - (1-z) \right) \frac{ F(Q^2,p \cdot Q, \bar{p} \cdot Q)}{Q^2}, \label{eq:dasf} \\
    F(Q^2,p\cdot Q, \bar{p} \cdot Q)&= \int \frac{\dd[d]k_1}{(2 \pi)^{d-1}} \int \frac{\dd[d]k_2}{(2 \pi)^{d-1}} \ \frac{\delta^+\left( k_1^2\right ) \delta^+\left( k_2^2\right )}{(p-k_1)^2 \ \bar{p}\cdot k_2 } \delta^d(Q-k_1-k_2). \label{eq:M4int}
 \end{align}
We first compute the function $F$ in \cref{eq:M4int} in the rest frame of the time-like vector $Q$. In that frame  \( Q=(Q_0,0,0,0) \) and $F$ becomes 
\begin{align}
    \begin{split}
  F&=-\frac{1}{2}\int \frac{\dd[d-1]\vec{k}_1}{(2 \pi)^{d-1}2|\vec{k}_1|} \int \frac{\dd[d-1]\vec{k}_2}{(2 \pi)^{d-1}2|\vec{k}_2|}\ \frac{\delta^{d-1}(\vec{k}_1+\vec{k}_2)}{\bar{p}_0 |\vec{k}_2|-\vec{\bar{p}} \cdot \vec{k}_2} \ \frac{\delta(Q_0-|\vec{k}_1|-|\vec{k}_2|) }{p_0 |\vec{k}_1|-\vec{p} \cdot \vec{k}_1 } \\
&=-\frac{1}{2}\int\frac{\dd[d-1]\vec{k}_1}{(2\pi)^{2d-2} 4 |\vec{k}_1|^2} \frac{\delta(Q_0-2|\vec{k}_1|)}{\bar{p}_0 |\vec{k}_1|+\vec{\bar{p}} \cdot \vec{k}_1 }\frac{1 }{p_0 |\vec{k}_1|-\vec{p} \cdot \vec{k}_1 }. \label{eq:howtoparaph}
    \end{split}
\end{align}
We parameterize the two light-like momenta as \(p= p_0 (1,\vec{n}_p) \), \(\bar{p}= \bar{p}_0 (1,\vec{n}_{\bar{p}}) \),\footnote{Note that in the rest frame of $Q$, $p$ and $\bar{p}$ are not in a back-to-back configuration.}
and introduce spherical coordinates for $\vec{k}_1$. We obtain
\begin{align}
    \begin{split}
F&= -\frac{1}{8 p_0 \bar{p}_0} \int \frac{\dd[d-1]\vec{k}_1}{(2 \pi)^{2d-2}|\vec{k}_1|^4} \delta\left(Q_0 - 2 |\vec{k}_1| \right)  \frac{1}{1-\vec{n}_p \cdot \vec{n}_k} \; \frac{1}{1+\vec{n}_{\bar{p}} \cdot \vec{n}_k}  \\
&= - \frac{1}{(2p_0 Q_0)(2\bar{p}_0 Q_0)} \left(\frac{Q_0}{2} \right)^{d-4} \int \frac{\dd[]\Omega_k^{(d-1)}}{(2 \pi)^{2d-2}} 
\frac{1}{(k_n \cdot p_1) \ (k_n \cdot p_2)} , \label{eq:M4int2}
    \end{split}
\end{align}
where we introduced the notation \( \ p_1=(1, \vec{n}_p) \), \( \ p_2=(1, -\vec{n}_{\bar{p}}) \) and \( \ k_n=(1,\vec{n}_{k}) \).
The angular integral in \cref{eq:M4int2} was discussed in Refs.~\cite{vanNeerven:1985xr,Somogyi:2011ir}. The result reads 
\begin{align}
\int 
\frac{\dd[]\Omega_{k}^{(d-1)}}{(k_n \cdot p_1) \ (k_n \cdot p_2)}&=-\Omega^{(d-2)} \frac{2^{-2\epsilon}}{\epsilon} \frac{\Gamma(1-\epsilon)^2}{\Gamma(1-2\epsilon)} \ _2F_1\left( 1,1,1-\epsilon,1-\frac{\rho_{12}}{2}\right), 
\end{align}
where  
$
\Omega^{(d)}= 2 \pi^{\frac{d}{2}}/ \Gamma(\tfrac{d}{2})
$
is the $d$-dimensional solid angle, $_2F_1$ is the Gauss hypergeometric function and $\rho_{12}=(1-\vec{n}_{p_1}\cdot\vec{n}_{p_2})$.
Finally, we rewrite $\rho_{12}$ in a Lorentz-invariant way
\begin{align}
    \frac{1-\rho_{12}}{2}&=\frac{1}{2}\left(1+\vec{n}_{p_1} \cdot \vec{n}_{p_2}\right)=\frac{Q^2}{(2 Q \cdot p) (2 Q \cdot \bar{p})}.
\end{align} 
The function $F$ in \cref{eq:M4int2} becomes
\begin{align}
    \begin{split}
    F(Q^2,p \cdot Q, \bar{p} \cdot Q) &=   \frac{\Omega^{(d-2)}}{(2 \pi)^{2d-2}} \frac{(Q^2)^{-\epsilon}}{(2 p \cdot Q)(2 \bar{p} \cdot Q)} \frac{\Gamma(1-\epsilon)^2 }{\epsilon \ \Gamma(1-2\epsilon)}  \\ 
    &\eqspace \times \ _2F_1 \left(1,1,1-\epsilon,\frac{Q^2}{(2 Q \cdot p) (2 Q \cdot \bar{p})}\right). \label{eq:Ftempin4}
    \end{split}
\end{align}
We substitute \cref{eq:Ftempin4} into \cref{eq:dasf} and find
\begin{align}
    \begin{split}
    I_6 \left(1,1,z\right)&= \int \dd[d]Q \ \delta\left( 2 Q \cdot p - \tfrac{1}{z} \right) \delta \left( 2 Q \cdot \bar{p} - (1-z) \right) \frac{\Omega^{(d-2)}}{(2 \pi)^{2d-2}} \frac{(Q^2)^{-1-\epsilon}}{(2 p \cdot Q)(2 \bar{p} \cdot Q)}  \\ 
    &\eqspace \times \frac{\Gamma(1-\epsilon)^2 }{\epsilon  \ \Gamma(1-2\epsilon)} \ _2F_1 \left(1,1,1-\epsilon,\frac{Q^2}{(2 Q \cdot p) (2 Q \cdot \bar{p})}\right).
    \end{split}
\end{align}
To integrate over $Q$ we employ the Sudakov decomposition $Q^\mu=\alpha p^\mu+\beta \bar{p}^{\mu}+Q_\perp^\mu$ so that
\begin{align}
    \begin{split}
    \int \dd[d]Q&=\frac{1}{2}\int_0^\infty \dd[]\alpha \int_0^\infty \dd[]\beta \int \dd[d-2] Q_{\perp}\\
    &=\frac{\Omega^{(d-2)}}{4}\int_0^\infty \dd[]\alpha \int_0^\infty \dd[]\beta \int\dd[]Q_\perp^2 \ (Q_\perp^2)^{-\epsilon} \ \theta\left( \alpha \beta-Q_\perp^2 \right).\label{eq:sud}
    \end{split}
\end{align}
In \cref{eq:sud} we used the fact that $Q^2>0$, $Q\cdot p>0$ and $Q \cdot \bar{p}>0$ to constrain integrations over $\alpha$ and $\beta$.
After eliminating the delta functions $\delta\left(\beta-1/z\right)$ and $\delta\left(\alpha-(1-z)\right)$ by integrating over $\alpha$ and $\beta$,
we obtain
\begin{align}
    \begin{split}
    I_6 \left(1,1,z\right)&= \frac{z}{(1-z)} \frac{\left[\Omega^{(d-2)}\right]^{2}}{4(2 \pi)^{2d-2}}\frac{\Gamma(1-\epsilon)^2 }{\epsilon \ \Gamma(1-2\epsilon)} \\
    &\eqspace\times \int_0^{\frac{1-z}{z}} \dd[]Q_\perp^2 \ (Q_\perp^2)^{-\epsilon} \left(\tfrac{1-z}{z}-Q_\perp^2 \right)^{-(1+\epsilon)} \ _2F_1\left(1,1,1-\epsilon,1-\tfrac{Q_\perp^2 \ z}{1-z} \right).
    \end{split}
\end{align}
We substitute $Q_\perp^2=(1-z)(1-u)/z$, integrate over $u$ and find
\begin{align}
    I_6 \left(1,1,z\right)&= - \frac{\left[\Omega^{(d-2)}\right]^2}{4 (2 \pi)^{2d-2}} \left(\frac{1-z}{z} \right)^{-1-2\epsilon} \frac{\Gamma(1-\epsilon)^4}{\epsilon^2 \ \Gamma(1-2\epsilon)^2} \ _3F_2 \left(1,1,-\epsilon;1-2\epsilon,1-\epsilon,1 \right),
\end{align}
where $ \ _3F_2$ is the generalized hypergeometric function~\cite{abramowitz+stegun}. The expansion of the hypergeometric function in $\epsilon$ is easily obtained using the program HypExp~\cite{Huber:2005yg,HypExp}.

Our next example is the master integral
\begin{align}
    I_9 \left(1,1,z\right)&= \int \frac{\dd[d]k_1}{(2 \pi)^{d-1}} \int \frac{\dd[d]k_2}{(2 \pi)^{d-1}} \ \delta^+\left( k_1^2\right )  \delta^+\left( k_2^2\right )  \frac{\delta\left( 2 k_{12} \cdot p -  \tfrac{1}{z} \right) \ \delta \left( 2 k_{12} \cdot \bar{p} - (1-z) \right) } {(p-k_{12})^2\ \ (p-k_2)^2 \ (p-k_1)\cdot\bar{p}}. 
\end{align}
We again insert $1=\int \dd[d]Q \ \delta^d(k_1+k_2-Q) $ into the integrand and write the integral as
\begin{align}
    I_9 \left(1,1,z\right)&= \int \dd[d]Q \ \delta\left( 2 Q \cdot p - \tfrac{1}{z} \right) \delta \left( 2 Q \cdot \bar{p} - (1-z) \right) \frac{ F_9(Q^2,p \cdot Q, \bar{p} \cdot Q)}{(p-Q)^2},  \label{eq:I9}\\
     F_9(Q^2,p\cdot Q, \bar{p} \cdot Q)&=  \int \frac{\dd[d]k_1}{(2 \pi)^{d-1}} \int \frac{\dd[d]k_2}{(2 \pi)^{d-1}} \ \frac{\delta^+\left( k_1^2\right ) \delta^+\left( k_2^2\right )}{ \bar{p}\cdot (p-k_1) \ (p-k_2)^2  } \delta^d(Q-k_1-k_2). \label{eq:MErichtig} 
 \end{align}
 We compute the integral \cref{eq:MErichtig} in the rest frame of the vector $Q$. To this end, we parameterize the phase space as shown in \cref{eq:howtoparaph}, integrate over $\vec{k}_1$ to remove the delta function, introduce spherical coordinates
 for $\vec{k}_2$ and integrate over the absolute value of $\vec{k}_2$ to remove the remaining delta function. We obtain the angular integral
 \begin{align} 
        \begin{split}
    F_9&= - \left( \frac{Q_0}{2}\right)^{d-2} \frac{1}{Q_0^2  \ Q_0 p_0 \ Q_0 \bar{p}_0} \frac{1}{\lambda} \int \frac{\dd[]\Omega_k^{(d-1)}}{(2 \pi)^{2d-2}} \frac{1}{1-\frac{1}{\lambda}\vec{n}_{\bar{p}} \cdot \vec{n}_k} \frac{1}{1-\vec{n}_p \cdot \vec{n}_k}   \\
    &= - \left( \frac{Q_0}{2}\right)^{d-2} \frac{1}{Q_0^2  \ Q_0 p_0 \ Q_0 \bar{p}_0} \frac{1}{\lambda} \int \frac{\dd[]\Omega_k^{(d-1)}}{(2 \pi)^{2d-2}} 
    \frac{1}{(k_n \cdot p_1) \ (k_n \cdot p_2)} , \label{eq:M4int9}
    \end{split}    
\end{align}
    where we introduced the notation \( \ p_1=(1, \frac{1}{\lambda}\vec{n}_{\bar{p}} ) \), with $\lambda=1/(Q_0 \bar{p}_0)-1$,\( \ p_2=(1,\vec{n}_{p}) \) and \( \ k_n=(1,\vec{n}_{k}) \).
    The angular integration in \cref{eq:M4int9} was discussed in Ref.~\cite{Somogyi:2011ir}; the result reads
    \begin{align}
        \begin{split}
        \int \frac{\dd[]\Omega_k^{(d-1)} }{(k_n \cdot p_1) \ (k_n \cdot p_2)}&=-\frac{1}{\epsilon}\frac{2^{1-2 \epsilon} \pi^{1-\epsilon} \ \lambda \ \Gamma(1-\epsilon)}{(\lambda-\vec{n}_{p} \cdot \vec{n}_{\bar{p}} ) \  \Gamma(1-2 \epsilon)} \\
         &\eqspace \times  F_1\left(1,-\epsilon,-\epsilon,1- 2 \epsilon,- \frac{1+\vec{n}_{p} \cdot \vec{n}_{\bar{p}}}{\lambda - \vec{n}_{p} \cdot \vec{n}_{\bar{p}}},\frac{-1+\vec{n}_{p} \cdot \vec{n}_{\bar{p}}}{-\lambda+\vec{n}_{p} \cdot \vec{n}_{\bar{p}}} \right). \label{eq:fApp}
    \end{split}
        \end{align}
    In \cref{eq:fApp} $ F_1$ is the Appell hypergeometric function (see e.g.~Ref.~\cite{gradshteyn2007}). Writing \cref{eq:fApp} in a Lorentz-invariant way, we obtain
    \begin{align}
        \begin{split}
        F(Q^2,p\cdot Q,& \bar{p} \cdot Q)=\frac{1}{(2 \pi)^{2d-2}} \frac{1}{\epsilon} \frac{\pi^{1-\epsilon} \ (Q^2)^{-\epsilon} \ \Gamma(1-\epsilon)}{(Q^2+2 p\cdot Q (1- 2\bar{p}\cdot Q))\Gamma(1-2\epsilon)}  \\
        &\times  F_1\left(1,-\epsilon,-\epsilon,1-2\epsilon,\frac{Q^2- 4 p\cdot Q \ \bar{p}\cdot Q }{Q^2+2 p\cdot Q (1-2 \bar{p}\cdot Q)},\frac{Q^2}{Q^2+2 p\cdot Q (1- 2 \bar{p}\cdot Q)}\right) \label{eq:appel}.
        \end{split}
    \end{align}
    We substitute \cref{eq:appel} into \cref{eq:I9}, introduce the Sudakov decomposition $Q^\mu=\alpha p^\mu+\beta \bar{p}^{\mu}+Q_\perp^\mu$ and integrate over $Q$ . 
    We substitute $Q_\perp^2=l (1-z)/z$ and find
    \begin{align}
            \begin{split}
        I_9 \left(1,1,z\right)&=-\frac{\Omega^{(d-2)}}{4 (2 \pi)^{2d-2}}\frac{1}{\epsilon} \int_0^1 \dd[]l \frac{ \pi^{1-\epsilon}\ z\ (1-z) \ }{[1-l\ (1-z)] \ [l\ (1-z)+z] \ } \left(\frac{(1-l) \ l \ (1-z)^2}{z^2}\right)^{-\epsilon} \\
        &\eqspace \times \frac{\Gamma(1-\epsilon)}{\Gamma(1- 2\epsilon)} F_1\left(1,-\epsilon,-\epsilon,1-2\epsilon,\frac{l \ (1-z) }{l(1-z)-1},\frac{(l-1) \ (1-z)}{l(1-z)-1}\right). \label{eq:thisappel}
            \end{split}
    \end{align}
    To perform the $l$-integration we use the integral representation of the Appell function~\cite{abramowitz+stegun}
    \begin{align}
        F_1\left(a,b_1,b_2,c,z_1,z_2 \right)= \int_0^1 \dd[]u \frac{\Gamma(c) \ u^{a-1} \ (1-u)^{c-a-1}}{\Gamma(a) \Gamma(c-a)} \ (1-u \ z_1)^{-b_1} (1- u\ z_2)^{-b_2}.
    \end{align}
     We find
    \begin{align}
        \begin{split}
        I_9 \left(1,1,z\right)&=  \int_0^1 \dd[]u \int_0^1 \dd[]l  \ \frac{\left[\Omega^{(d-2)}\right]^2}{(2\pi)^{2d-2}}\frac{(z-1)\ z \ (1-u)^{-1-2\epsilon} \ \Gamma(1-\epsilon)^2}{4 \ [1+l \ (z-1)]\ [l\ (z-1)-z] \ \Gamma(1-2\epsilon)} \\
        &\eqspace\times \left[ \frac{(1-l) \ l \ (1-z)^2}{z} \right]^{-\epsilon} \left[ \frac{z\ [1+l \ (-1+u+z-u \ z)]}{1+l\ (z-1)}\right]^{\epsilon} \\
        &\eqspace\times\left[\frac{1+u\ (z-1)+ l \ (-1+u+z-u z)}{1+l\ (z-1)} \right]^\epsilon. \label{eq:doubleint}
        \end{split}
    \end{align}

    We would like to expand the integrand in a Laurent series in $\epsilon$
and compute the integral order by order in this expansion. This can be done if the integrand remains integrable at $\epsilon=0$.
It is easy to see that this is not the case;
while the integral over $l$ in \cref{eq:doubleint} converges if we Taylor expand around $\epsilon=0$, the integral over $u$ diverges at $u=1$.

We remove the divergence by performing an end-point subtraction at $u=1$, splitting the integral into two pieces.  
    To write the result, we define two functions
    \begin{align}
        M(u,l)&=\left[ \frac{z\ [1+l \ (-1+u+z-u \ z)]}{1+l\ (z-1)}\right]^{\epsilon}\left[\frac{1+u\ (z-1)+ l \ (-1+u+z-u z)}{1+l\ (z-1)} \right]^\epsilon,\\
        G(l)&=\frac{\left[\Omega^{(d-2)}\right]^2}{(2\pi)^{2d-2}}\frac{(z-1)\ z  \ \Gamma(1-\epsilon)^2}{4 \ [1+l \ (z-1)]\ [l\ (z-1)-z] \ \Gamma(1-2\epsilon)} \left[ \frac{(1-l) \ l \ (1-z)^2}{z} \right]^{-\epsilon},
    \end{align}
   and write \cref{eq:doubleint} as
    \begin{align}
    \begin{split}
        I_9 \left(1,1,z\right)&=  \int_0^1 \dd[]u \int_0^1 \dd[]l   \ (1-u)^{-1-2\epsilon} \ G(l) \ M(u,l) \\
    &=  \int_0^1 \dd[]u \int_0^1 \dd[]l \ (1-u)^{-1-2\epsilon} \ G(l) \ \left[M(u,l)-M(1,l) \right]  \\
    &\eqspace+\int_0^1 \dd[]u \int_0^1 \dd[]l \ (1-u)^{-1-2\epsilon} \ G(l) \  M(1,l). \label{eq:splitint}
    \end{split}
    \end{align}
    The $u=1$ singularity in the first term on the right hand side of \cref{eq:splitint} is now regulated, while the last term in \cref{eq:splitint} can be easily integrated over $u$. We find
    \begin{align}
        \begin{split}
         I_9 \left(1,1,z\right)
        &=  \int_0^1 \dd[]u \int_0^1 \dd[]l \ (1-u)^{-1-2\epsilon} \ G(l) \ \left[M(u,l)-M(1,l) \right] \\
        &\eqspace- \frac{1}{2 \epsilon}\int_0^1 \dd[]l  \ G(l) \  M(1,l). \label{eq:splitint2}
        \end{split}
        \end{align}
All remaining integrands in \cref{eq:splitint2} can now expanded to the required order in $\epsilon$ and integrated using the HyperInt package~\cite{Panzer:2014caa}.
The final result reads
\begin{align}
    \begin{split}
    I_9 \left(1,1,z\right)&=\frac{\left[\Omega^{(d-2)}\right]^2}{(2\pi)^{2d-2}} \ (1-z)^{-2\epsilon}\Bigg[\frac{1}{\epsilon}\;\frac{z } { \ 4 \ (1+z)} \ H(0,z) \\
    &\eqspace- \frac{z}{8\ (1+z)}\left( \pi^2+4\ H(-1,0,z)-8 \ H(0,0,z)+4 \ H(1,0,z)  \right)\Bigg]+ \order{\epsilon}
    \end{split}
\end{align}
where  $H(\vec{m}_w,z)$
are harmonic polylogarithms (HPLs)~\cite{HPLs}.

% \subsubsection{Example master integral $\tilde{I}[10]$}
Finally, we consider the real-virtual master integral $I_{10}$. It reads 
\begin{align}
    I_{10} \left(1,1,z\right)& =\int \frac{\dd[d]k}{(2 \pi)^{d-1}}\int \frac{\dd[d]l}{(2 \pi)^{d}} \ \delta^+\left( k^2\right ) \  \frac{\delta\left( 2 k \cdot p - \tfrac{1}{z} \right) \ \delta\left( 2 k \cdot \bar{p} - (1-z) \right)}{l^2 \  (l \cdot \bar{p})  \ (p-l)^2  \  (p-k-l)^2} \label{eq:I10} . 
\end{align}
We perform the $l$-integration first.
 To this end, we combine the propagators \(1 / l^2 \) and \(1 / l \cdot \bar{p} \ \). We write
\begin{align}
  \frac{1}{l^2} \frac{1}{(2 l \cdot \bar{p})} = \int_0^\infty \frac{\dd[]y}{(l^2+2  \ l \cdot \bar{p}  \ y)^2} 
  =\int_0^\infty \frac{\dd[]y}{[(l+y   \ \bar{p} )^2]^2} \label{eq:linjoined},
\end{align}
and obtain the standard loop integral over $l$
\begin{align}
 \int_0^\infty \dd[]y \int \frac{\dd[d]l}{(2 \pi)^d}  \frac{1}{[(l+y  \  \bar{p} )^2]^2  \ (p-l)^2  \ (p-k-l)^2}.
\end{align}
 The integration is now straightforward and we obtain
\begin{align}
    \begin{split}
  \int \frac{\dd[d]l}{(2 \pi)^d}  \frac{1}{l^2 \ (l \cdot \bar{p})  \ (p-l)^2  \ (p-k-l)^2} &= -i  \ 2^{-2+2 \epsilon} \pi^{-2+\epsilon} \frac{  \Gamma(1-\epsilon)^2 \Gamma(1+\epsilon)}{\epsilon^2 \ \Gamma(1- 2 \epsilon)} \\
  &\eqspace \times (2 p \cdot k)^{-1-\epsilon} \ _2  F_1\left( 1, -\epsilon, 1-\epsilon, 2 \bar{p} \cdot k \right).
    \end{split}
\end{align}
The remaining integration over the on-shell momentum $k$ is performed by introducing the Sudakov decomposition $k^\mu=\alpha p^\mu +\beta \bar{p}^\mu+k_\perp^\mu$. We find
\begin{align}
    I_{10} \left(1,1,z\right) &=-i \frac{ \left[\Omega^{(d-2)}\right]^2}{4  (2 \pi)^{2d-2}} \ (1-z)^{-\epsilon} z^{1+2\epsilon} \frac{\Gamma(1-\epsilon)^3 \Gamma(1+\epsilon)}{\epsilon^2 \ \Gamma(1- 2 \epsilon)} \ _2F_1(1,-\epsilon,1-\epsilon,1-z) .
\end{align}

This concludes the discussion of the evaluation of the master integrals. All manipulations with hypergeometric functions that appear in master integrals,
including their expansions in $\epsilon$,
are performed with the help of the HypExp package~\cite{HypExp}. We describe some results for the beam functions in the next section.
\subsection{Results}
\label{subsec:resultsbeam}
We are now in a position to present the bare partonic beam functions
 $B^{b}_{q_iq_j}$, $B^b_{q_ig}$, $B^b_{q_i\bar{q}_j}$, $B^b_{gg}$ and $B^b_{gq_i}$ through $\order{\epsilon^2}$ at NNLO QCD.
 By performing the renormalization procedure and matching onto
 partonic distribution functions, as discussed in Refs.~\cite{Stewart:2009yx,Tackmann2014,Gaunt:2014cfa,Behring:2019quf}, we also obtain the matching coefficients
 $I_{q_iq_j}$, $I_{q_ig}$, $I_{q_i\bar{q}_j}$, $I_{gg}$ and $I_{gq_i}$.
 To present the results, we write the beam functions and the matching coefficients as a series in the renormalized $\overline{\mathrm{MS}}$ coupling constant
 \begin{align}
    B^{b}_{ij}&=\sum_{k=0}^n \left(\frac{\alpha_s}{4 \pi}\right)^k B^{b \ (k)}_{ij},
    &I_{ij}&=\sum_{k=0}^n \left(\frac{\alpha_s}{4 \pi}\right)^k I^{(k)}_{ij}. 
 \end{align}
 Since the expressions for the bare partonic beam functions $B^b_{ij}$ and the matching coefficients $I_{ij}$ through $\order{\epsilon^2}$ are lengthy,
 we only discuss some features of the most complicated coefficient $I_{gg}$; complete expressions for all other matching coefficients are given in an ancillary file provided with this submission.
 We write the matching coefficient in the following form
 \begin{align}
    I^{(2)}_{gg}&=\sum_{k=0}^5 \frac{1}{\mu^2}L_k\left(\frac{t}{\mu^2} \right) F_+^{(k)}(z)+\delta(t) F_\delta(z), \\
    F_\delta(z)&=C_{-1} \delta(1-z)+ \sum_{k=0}^5 C_k L_k(1-z)+F_{\delta,h}(z),
 \end{align}
 where we define the plus distribution
 \begin{align}
    L_n(z)=\left[\frac{\ln^n(z)}{z} \right]_+.
\end{align}
 For brevity, we only show the coefficient $C_{-1}$ as well as the function $F_{\delta,h}(z)$ in pure gluodynamics ($n_f=0$). 
 For the coefficient $C_{-1}$ we find
 \begin{align}
    \begin{split}
    C_{-1}&= \  C_A^2 \left(-\frac{110 \zeta (3)}{9}+\frac{2428}{81}-\frac{67 \pi ^2}{18}+\frac{11 \pi ^4}{90}\right)+C_A n_f T_F \left(\frac{40 \zeta (3)}{9}-\frac{656}{81}+\frac{10 \pi
    ^2}{9}\right)  \\
    &\eqspace+ \epsilon \Bigg[C_A^2 \left(-\frac{938 \zeta (3)}{27}+\frac{65 \pi ^2 \zeta (3)}{3}-150 \zeta (5)+\frac{14576}{243}-\frac{202 \pi ^2}{27}+\frac{77 \pi ^4}{540}\right) \\
    &\eqspace+C_A n_f T_F
    \left(\frac{280 \zeta (3)}{27}-\frac{3904}{243}+\frac{56 \pi ^2}{27}-\frac{7 \pi ^4}{135}\right) \Bigg] \\
    &\eqspace + \epsilon^2 \Bigg[ C_A^2 \Bigg(-\frac{5656 \zeta (3)}{81}+\frac{220 \pi ^2 \zeta (3)}{27}+\frac{1142 \zeta (3)^2}{9}-\frac{638 \zeta (5)}{15}+\frac{87472}{729} \\
    &\eqspace-\frac{1214 \pi ^2}{81}+\frac{67 \pi
    ^4}{216}-\frac{593 \pi ^6}{11340}\Bigg)+C_A n_f T_F \Bigg(\frac{1568 \zeta (3)}{81}-\frac{80 \pi ^2 \zeta (3)}{27}+\frac{232 \zeta (5)}{15} \\
    &\eqspace-\frac{23360}{729}+\frac{328 \pi
    ^2}{81}-\frac{5 \pi ^4}{54}\Bigg)\Bigg].
    \end{split}
 \end{align}
 To present the result for the function $F_{\delta,h}(z)$ in gluodynamics we write
 \begin{align}
  F_{\delta,h}(z)|_{n_f=0}&=C_A^2 \left( F_0(z)+ \epsilon  \ F_1(z)+ \epsilon^2 \ F_2(z) \right),
 \end{align}
 and introduce the short-hand notation $H_{\vec{a}}=H(\vec{a},z)$.
 Due to its large size, we do not display the function $F_2$ and only show the functions $F_0$ and $F_1$.
 They read
 \begin{footnotesize}
 \begin{align}
    F_0&=48 \left(z^2-z-\frac{1}{z}+2\right) H_{1,1,1}+\frac{4 \left(55 z^3-47 z^2+58 z-55\right) H_{1,1}}{3 z}\notag\\
    &\eqspace+\frac{2 \left(286 z^4-365 z^3+342 z^2-307 z+66\right) H_{0,0}}{3 (z-1) z}+\frac{4 \left(55
    z^4-102 z^3+105 z^2-102 z+55\right) H_{1,0}}{3 (z-1) z}\notag\\
    &\eqspace+\frac{32 \left(z^4-3 z^3+3 z^2-z+1\right) H_{2,0}}{(z-1) z}+\frac{8 \left(7 z^4-18 z^3+21 z^2-10 z+7\right) H_{2,1}}{(z-1) z}\notag\\
    &\eqspace+\frac{8
    \left(3 z^4-10 z^3-7 z^2+10 z+7\right) H_{0,0,0}}{(z-1) (z+1)}+\frac{8 \left(6 z^4-12 z^3+18 z^2-11 z+6\right) H_{1,1,0}}{(z-1) z}\notag\\
    &\eqspace+\frac{\left(z^2+z+1\right)^2}{z (z+1)} \left(-16 H_{-2,0}-16 H_{-1,2}+16 H_{-1,-1,0}-32
    H_{-1,0,0}+4 \pi ^2 H_{-1}\right)\\
    &\eqspace+\frac{\left(z^2-z+1\right)^2}{(z-1) z} \left(56 H_{1,2}+56 H_{1,0,0}\right)+\frac{16 H_3 \left(4 z^5-7 z^4+7 z^2+3\right)}{z \left(z^2-1\right)}\notag\\
    &\eqspace+H_1 \left(\frac{2 \left(134 z^4+102 z^3+131
    z^2+163 z-134\right)}{9 z (z+1)}-\frac{4 \pi ^2 \left(7 z^4+7 z^2+13 z-7\right)}{3 z (z+1)}\right)\notag\\
    &\eqspace+\frac{4 H_2 \left(99 z^4-133 z^3+123 z^2-111 z+33\right)}{3 (z-1) z}+H_0 \Bigg(\frac{-268 z^4-563
    z^3+462 z^2-167 z+804}{9 (z-1) z}\notag\\
    &\eqspace-\frac{\pi ^2 \left(44 z^5-60 z^4+12 z^3+64 z^2-8 z+28\right)}{3 (z-1) z (z+1)}\Bigg)-\frac{2 \pi ^2 \left(99 z^4+65 z^3+55 z^2+67 z-33\right)}{9 z (z+1)}\notag\\
    &\eqspace+\frac{2
    \left(2460 z^4+553 z^3+350 z^2+255 z-2406\right)}{27 z (z+1)}-\frac{\left(120 z^5-112 z^4+88 z^3+120 z^2-200 z+80\right) \zeta (3)}{(z-1) z (z+1)}, \notag \\
    F_1&=-\frac{4 \zeta (3) \left(154 z^4+97 z^3+96 z^2+109 z-66\right)}{3 z (z+1)}+\frac{\pi ^2 \left(402 z^4+2323 z^3+2618 z^2+2037 z+1742\right)}{54 z (z+1)}\notag\\
    &\eqspace+\frac{16627 z^4+12881 z^3+4460 z^2-3169
    z-16231}{81 z (z+1)}+\frac{\pi ^4 \left(15 z^5-83 z^4-41 z^3+83 z^2+11 z+27\right)}{45 (z-1) z (z+1)}\notag\\
    &\eqspace+\left(-\frac{\pi ^2 \left(64 z^4-168 z^3+192 z^2-84 z+64\right)}{3 (z-1) z}-\frac{2 \left(134
    z^4+841 z^3-708 z^2+403 z-938\right)}{9 (z-1) z}\right) H_2\notag\\
    &\eqspace+\frac{4 \left(341 z^4-443 z^3+429 z^2-393 z+99\right) H_3}{3 (z-1) z}+\frac{16 \left(10 z^5-19 z^4-2 z^3+19 z^2+2 z+7\right) H_4}{z
    \left(z^2-1\right)}\notag\\
    &\eqspace+\left(\frac{-1340 z^4-699 z^3+66 z^2+633 z+1876}{9 (z-1) z}-\frac{4 \pi ^2 \left(27 z^5-38 z^4+7 z^3+38 z^2-7 z+18\right)}{3 z \left(z^2-1\right)}\right) H_{0,0}\notag\\
    &\eqspace+\left(\frac{4
    \left(201 z^4-302 z^3+333 z^2-299 z+201\right)}{9 (z-1) z}-\frac{4 \pi ^2 \left(22 z^4-44 z^3+66 z^2-43 z+22\right)}{3 (z-1) z}\right) H_{1,0}\notag\\
    &\eqspace+\left(\frac{2 \left(402 z^4+197 z^3+259 z^2+330
    z-402\right)}{9 z (z+1)}-\frac{8 \pi ^2 \left(10 z^4+10 z^2+19 z-10\right)}{3 z (z+1)}\right) H_{1,1}\notag\\
    &\eqspace+\frac{4 \left(55 z^4-102 z^3+105 z^2-102 z+55\right) H_{1,2}}{(z-1) z}+\frac{4 \left(253
    z^4-344 z^3+333 z^2-308 z+99\right) H_{2,0}}{3 (z-1) z}\notag\\
    &\eqspace+\frac{4 \left(253 z^4-344 z^3+333 z^2-308 z+99\right) H_{2,1}}{3 (z-1) z}+\frac{32 \left(4 z^4-11 z^3+12 z^2-5 z+4\right) H_{2,2}}{(z-1)
    z}\notag\\
    &\eqspace+\frac{16 \left(7 z^5-15 z^4-3 z^3+15 z^2+3 z+5\right) H_{3,0}}{z \left(z^2-1\right)}+\frac{32 \left(6 z^5-10 z^4+z^3+10 z^2-z+5\right) H_{3,1}}{z \left(z^2-1\right)}\notag\\
    &\eqspace+\frac{2 \left(770 z^4-985
    z^3+954 z^2-871 z+198\right) H_{0,0,0}}{3 (z-1) z}+\frac{4 \left(121 z^4-214 z^3+219 z^2-214 z+121\right) H_{1,0,0}}{3 (z-1) z}\notag \\
    &\eqspace+\frac{4 \left(55 z^4-102 z^3+105 z^2-102 z+55\right)
    H_{1,1,0}}{(z-1) z}+4 \left(55 z^2-47 z+58-\frac{55}{z}\right) H_{1,1,1}\notag\\
    &\eqspace+\frac{8 \left(20 z^4-40 z^3+60 z^2-39 z+20\right) H_{1,1,2}}{(z-1) z}+\frac{16 \left(5 z^5-9 z^4-z^3+9 z^2+z+3\right)
    H_{2,0,0}}{z \left(z^2-1\right)}\notag\\
    &\eqspace+\frac{48 \left(3 z^4-8 z^3+9 z^2-4 z+3\right) H_{2,1,0}}{(z-1) z}+\frac{24 \left(7 z^4-18 z^3+21 z^2-10 z+7\right) H_{2,1,1}}{(z-1) z}\notag\\
    &\eqspace+\frac{8 \left(3 z^5-22
    z^4-23 z^3+22 z^2+23 z-4\right) H_{0,0,0,0}}{z \left(z^2-1\right)}\notag\\
    &\eqspace+\frac{\left(z^2-z+1\right)^2}{(z-1) z} \left(176 H_{1,3}+144 H_{1,2,0}+168 H_{1,2,1}+136 H_{1,0,0,0}+112 H_{1,1,0,0}\right)\notag\\
    &\eqspace+\frac{24 \left(6 z^4-12 z^3+18
    z^2-11 z+6\right) H_{1,1,1,0}}{(z-1) z}+\frac{160 \left(z^3-z^2+2 z-1\right) H_{1,1,1,1}}{z}\notag\\
    &\eqspace+H_1 \Bigg(-\frac{5 \pi ^2 \left(55 z^4+8 z^3+11 z^2+14 z-55\right)}{9 z (z+1)}+\frac{2 \left(5333
    z^4+1111 z^3+721 z^2+472 z-5279\right)}{27 z (z+1)}\notag\\
    &\eqspace-\frac{8 \left(56 z^4+56 z^2+115 z-56\right) \zeta (3)}{3 z (z+1)}\Bigg)\notag\\
    &\eqspace+H_0 \Bigg(\frac{\pi ^2 \left(583 z^4-753 z^3+735 z^2-675
    z+165\right)}{9 (1-z) z}+\frac{2 \left(7021 z^4-10345 z^3+9138 z^2-8624 z+3609\right)}{27 (z-1) z}\\
    &\eqspace-\frac{16 \left(49 z^5-49 z^4+31 z^3+53 z^2-27 z+31\right) \zeta (3)}{3 z
    \left(z^2-1\right)}\Bigg)\notag\\
    &\eqspace+\frac{\left(z^2+z+1\right)^2}{z (z+1)} \Bigg(8 \pi ^2 H_{-2}-32 H_{-3,0}-32 H_{-2,2}-8 \pi ^2 H_{-1,-1}+12 \pi ^2 H_{-1,0}-64 H_{-1,3}+32 H_{-2,-1,0}\notag\\
    &\eqspace-64 H_{-2,0,0}+32 H_{-1,-2,0}+32
    H_{-1,-1,2}-32 H_{-1,2,0}-32 H_{-1,2,1}-32 H_{-1,-1,-1,0}+64 H_{-1,-1,0,0}\notag\\
    &\eqspace-80 H_{-1,0,0,0}+56 H_{-1} \zeta (3)\Bigg) \notag.
\end{align}
\end{footnotesize}%

Computer-readable expressions for all partonic beam functions and matching coefficients can be found in an ancillary file provided with this submission.
We check the results for all matching coefficients
against the $\order{\epsilon^0}$ results in Refs.~\cite{Tackmann2014,Gaunt:2014cfa} and find full agreement. We discuss the calculation of the soft function in the next section.
% -------------------------
\section{Calculation of the soft function}
\label{sec:soft}
In this section we describe the calculation of the bare zero-jettiness soft function $S$ at NNLO in QCD. We begin by discussing the general setup in \cref{sec:softsetup},
 relating the calculation of the soft function to soft limits of QCD amplitudes for color singlet production. We re-write step functions, that originate from the zero-jettiness measure, as integrals of delta functions over auxillary
parameters. We then use reverse
unitarity to express the soft function through master integrals.
In \cref{sec:softmasters} we describe the calculation of master integrals as functions of the auxillary parameters and explain in \cref{sec:zint} how the
remaining integrations over auxillary parameters can be performed.
\subsection{General setup}
\label{sec:softsetup}
The zero-jettiness bare soft function can be calculated by considering soft limits of scattering amplitudes for colour singlet production.
These soft limits, described by eikonal functions, were calculated through NNLO QCD in Refs.~\cite{Catani:1999ss,Catani:2000pi}. We extract them from that reference and integrate the obtained expression
over the $m$-particle unresolved phase space $\dd[]\textrm{PS}_S^{(m)}$  including the $m$-particle zero-jettiness measure $M_m$ for the set of radiated partons $\{m\}$ with momenta $k_m$.

We begin by writing the bare soft function as a series in the bare strong coupling constant
\begin{align}
    S&=\sum_{i=0}^n [\alpha_s]^i S^{(i)}, 
\end{align}
where we defined
\begin{align}
    [\alpha_s]=\frac{g_{b,s}^2}{8 
\pi^2} \frac{(4 \pi)^\epsilon}{\Gamma(1-\epsilon)}.
\end{align}
The lower order results read
\begin{align}
S^{(0)}&=\delta(\tau), 
&S^{(1)}&=4 \ C_a \frac{\tau^{-1-2 \epsilon}}{\epsilon},
\end{align}
where $C_a=C_F(C_A)$ if the incoming particles are quarks(gluons), respectively.
At NNLO we need to consider the following contributions to the soft function
\begin{align}
    \begin{split}
   S^{(2)}&= \int \dd[]\textrm{PS}_S^{(1)} \ M_1 \ \xi^{(2)}_{g}+\frac{1}{2!}\int \dd[]\textrm{PS}_S^{(2)} \ M_2 \ \xi^{(2)}_{gg}+\int \dd[]\textrm{PS}_S^{(2)} \ M_2 \ \xi^{(2)}_{q\bar{q}},\label{eq:NNLOsoft}\\
           &= S^{(2)}_{g}+S^{(2)}_{gg}+S^{(2)}_{q\bar{q}},
    \end{split}
\end{align} 
where the functions $\xi^{(2)}_{g,q\bar{q},gg}$ denote various eikonal functions and for $m=1,2$ we introduced the short-hand notation
\begin{align}
    \dd[]\mathrm{PS}_S^{(m)}=\left( \frac{8 \pi^2 \ \Gamma(1-\epsilon)}{(4 \pi)^\epsilon} \right)^{2}\prod_n^m \frac{\dd[d]{k_n}}{{(2 \pi)}^{d-1}}  \  \delta^+ \left( k_n^2 \right) . \label{eq:PSSm}
\end{align} 
The first term in \cref{eq:NNLOsoft} describes the emission of one real gluon and an additional loop correction. The second and third terms in \cref{eq:NNLOsoft} describe the emission of two gluons and the emission of a quark anti-quark pair, respectively.
The single gluon emission contribution  $S^{(2)}_{g}$ has been calculated to arbitrary order in $\epsilon$ in Ref.~\cite{Monni:2011gb}. It reads
\begin{align}
    S^{(2)}_{g}=-2 C_a C_A \frac{ \Gamma (1-\epsilon)^5 \Gamma (1+\epsilon)^3}{\Gamma (1-2
    \epsilon)^2 \Gamma (1+2 \epsilon)} \frac{\tau ^{-1-4 \epsilon}}{\epsilon^3 },
\end{align}
and we thus focus on the double-real emission pieces.

The zero-jettiness measure for two real partons reads~\cite{Monni:2011gb}
\begin{align}
    \begin{split}
    % M
    M_2&= [ \delta\left( \tau - 2 p\cdot k_1 - 2 p \cdot k_2 \right) \theta\left( 2 \bar{p} \cdot k_1 - 2 p\cdot k_1 \right) \theta \left( 2 \bar{p} \cdot k_2 - 2 p \cdot k_2 \right)  + \left( p^{\mu} \leftrightarrow \bar{p}^{\mu}\right)   \\
    &\eqspace + \  \delta\left( \tau - 2 \bar{p}\cdot k_1 - 2 p \cdot k_2 \right) \theta\left( 2 p \cdot k_1 - 2 \bar{p}\cdot k_1 \right) \theta \left( 2 \bar{p} \cdot k_2 - 2 p \cdot k_2 \right)  + \left( p^{\mu} \leftrightarrow \bar{p}^{\mu}\right)], \label{eq:M2}
    \end{split}
\end{align}
where the momenta $p$ and $\bar{p}$ are again two complementary light-like vectors and we set $p \cdot \bar{p}=1/2$.
We refer to different sets of delta functions and step functions in \cref{eq:M2} as ``configurations''. Since the integrands in \cref{eq:NNLOsoft}
 are invariant under exchange of $p$ and $\bar{p}$,
  % a
 it is sufficient to only consider two configurations $M_2=2 \ M_A+2 \ M_B$, which we refer to as $A$ and $B$. Hence, we write
 \begin{align}
    M_A(k_1,k_2)&=\delta\left( \tau - 2 p\cdot k_1 - 2 p \cdot k_2 \right) \theta\left( 2 \bar{p} \cdot k_1 - 2 p\cdot k_1 \right) \theta \left( 2 \bar{p} \cdot k_2 - 2 p \cdot k_2 \right), \label{eq:MA} \\
    M_B(k_1,k_2)&=\delta\left( \tau - 2 \bar{p}\cdot k_1 - 2 p \cdot k_2 \right) \theta\left( 2 p \cdot k_1 - 2 \bar{p}\cdot k_1 \right)\theta \left( 2 \bar{p} \cdot k_2 - 2 p \cdot k_2 \right). \label{eq:MB}
 \end{align}
For color-singlet production, the quantities $\xi^{(2)}_{q\bar{q}}$ and  $\xi^{(2)}_{gg}$ in \cref{eq:NNLOsoft} can be found in Eq.~(A1) and Eq.~(A3) of Ref.~\cite{Catani:1999ss}
 \begin{align}
    \xi^{(2)}_{q\bar{q}}&=   \ T_F \ C_a \left(\mathcal T_{11} +\mathcal T_{22} -2 \mathcal T_{12} \right), \label{eq:eikq} \\
    \xi^{(2)}_{gg}&=    C_a \ \left[ 4 \ C_a \ \xi_{12}(k_1) \ \xi_{12}(k_2)+ C_A \left(2\xi_{12}-\xi_{11}-\xi_{22} \right) \right], \label{eq:eikg}
 \end{align} 
 where 
\begin{align}
    \mathcal T_{ij}&=-\frac{2 (p_i \cdot p_j)(k_1 \cdot k_2)+[p_i \cdot (k_1-k_2)][p_j\cdot (k_1-k_2)]}{2 (k_1 \cdot k_2)^2[p_i \cdot (k_1+k_2)][p_j \cdot (k_1+k_2)]}, \\
    \begin{split}
    \xi_{ij}&=\frac{(1-\epsilon)}{(k_1 \cdot k_2)^2}\frac{p_i \cdot k_1 \ p_j \cdot k_2 + p_j \cdot k_1 \ p_i \cdot k_2}{p_i \cdot (k_1 + k_2) \ p_j \cdot (k_1 + k_2)}  \\
    &\eqspace- \frac{(p_i \cdot p_j)^2}{2 p_i\cdot k_1 \ p_j \cdot k_2 \ p_i \cdot k_2 \ p_j \cdot k_1} \left[2-\frac{p_i\cdot k_1 \ p_j \cdot k_2+p_i\cdot k_2 \ p_j \cdot k_1 }{p_i \cdot (k_1 + k_2) \ p_j \cdot (k_1 + k_2)} \right] \\
    &\eqspace+\frac{p_i \cdot p_j}{2 k_1 \cdot k_2} \Bigg[\frac{2}{p_i \cdot k_1 \ p_j \cdot k_2} + \frac{2}{p_j \cdot k_1 \ p_i \cdot k_2}  \\
    &\eqspace- \frac{1}{p_i\cdot (k_1+k_2) p_j \cdot (k_1+k_2)} \left(4 + \frac{\left(p_i \cdot k_1 \ p_j \cdot k_2+ p_i \cdot k_2 \ p_j \cdot k_1  \right)^2}{p_i \cdot k_1 \ p_j \cdot k_2 \ p_i \cdot k_2 \ p_j\cdot k_1} \right) \Bigg], 
    \end{split}\\
    \xi_{ij}(k_1)&=\frac{p_i \cdot p_j}{(p_i\cdot k_1)(p_j \cdot k_1)},
\end{align}
with $p_1=p$, $p_2=\bar{p}$. 

% T
We note that $S^{(2)}_{q\bar{q}}$ and $S^{(2)}_{gg}$ were obtained in Refs.~\cite{Monni:2011gb,Kelley:2011ng} by directly integrating $\mathcal T_{ij}$ and $\xi_{ij}$ over the relevant phase space.
We will discuss an alternative to this approach,
 that is in line with the beam function calculation discussed in \cref{sec:calcset}. We hope that this approach can be extended to enable an N$^3$LO calculation of the zero-jettiness soft function.

 To this end, we would like to employ reverse unitarity and IBP technology to simplify calculation of the soft function. To do so, we map step functions 
  on to delta functions, using the following identity
% A
\begin{align}
    \theta(b-a)=&\int_0^1 \dd[]z \  \delta \left( z \ b  - a \right) \ b ,\label{eq:nicedelta}
\end{align}
which holds for $a,b \in [0,\infty)$. 
 Since $k_{1,2} \cdot p$, $k_{1,2} \cdot \bar{p}$ $\in [0,\infty)$, \cref{eq:nicedelta} is applicable. We therefore rewrite \cref{eq:MA,eq:MB}
as follows
\begin{align}
    \begin{split}
    M_A&=\delta\left( \tau - 2 p\cdot k_1 - 2 p \cdot k_2 \right) \theta\left( 2 \bar{p} \cdot k_1 - 2 p\cdot k_1 \right) \theta \left( 2 \bar{p} \cdot k_2 - 2 p \cdot k_2 \right)  \\
        &= \int_0^1 \dd[]z_1 \int_0^1 \dd[]z_2 \ \delta\left( \tau - 2 p\cdot k_1 - 2 p \cdot k_2 \right)  \delta\left( 2 z_1 \bar{p} \cdot k_1 - 2 p\cdot k_1 \right) 2 \bar{p} \cdot k_1  \\
        &\eqspace \times  \delta \left( 2 z_2 \bar{p} \cdot k_2 - 2 p \cdot k_2 \right) 2 \bar{p} \cdot k_2, \label{eq:M2A} 
    \end{split}\\
    \begin{split}
        M_B&=\delta\left( \tau - 2 \bar p\cdot k_1 - 2 p \cdot k_2 \right) \theta\left( 2 p \cdot k_1 - 2 \bar p\cdot k_1 \right) \theta \left( 2 \bar{p} \cdot k_2 - 2 p \cdot k_2 \right)  \\
        &= \int_0^1 \dd[]z_1 \int_0^1 \dd[]z_2 \ \delta\left( \tau - 2 \bar p\cdot k_1 - 2 p \cdot k_2 \right)  \delta\left( 2 z_1 p \cdot k_1 - 2 \bar p\cdot k_1 \right) 2 p \cdot k_1  \\
        &\eqspace \times \delta \left( 2 z_2 \bar{p} \cdot k_2 - 2 p \cdot k_2 \right) 2 \bar{p} \cdot k_2. \label{eq:M2B} 
    \end{split}
    \end{align}
\cref{eq:M2A,eq:M2B}, allow us to use reverse unitarity and IBP relations to express the soft function in terms of master integrals.

To illustrate this point, we discuss the computation of $S_{q\bar{q}}^{(2)}$ in detail; the computation of $S_{gg}^{(2)}$ is analogous.
According to our earlier discussion, contributions to the soft functions due to an emission of a $q\bar{q}$ pair read 
\begin{align}
    \begin{split}
    S^{(2)}_{q\bar{q}}&=2 \int \dd[]\textrm{PS}^{(2)}_S \ M_A \ \xi^{(2)}_{q\bar{q}}+2 \int \dd[]\textrm{PS}^{(2)}_S \ M_B \ \xi^{(2)}_{q\bar{q}} \\
    &= \ T_F \ C_a \ n_f \left(2 \ S^{(2)}_{q\bar{q},A}+2 \ S^{(2)}_{q\bar{q},B}\right). \label{eq:qqbarS}
    \end{split}
\end{align}
We note that we have split \cref{eq:qqbarS} into two contributions, stemming from configurations $A$ and $B$. They read
\begin{align}
    S^{(2)}_{q\bar{q}, A,B}= \int \dd[]\textrm{PS}^{(2)}_S\ M_{A,B} \left( \mathcal T_{11} +\mathcal T_{22} -2 \mathcal T_{12} \right).
\end{align}

We proceed by writing all delta functions in \cref{eq:M2A,eq:M2B} as linear combinations of the corresponding ``propagators'' and performing partial fractioning. We find that in configuration $A$ all integrals can be mapped onto two integral families 
\begin{align}
    \begin{split}
  I^{q\bar{q},1}_{n_1n_2}&= \Bigg\langle \left(p \cdot k_1 \right)^{-n_1} \left(\ k_1 \cdot k_2 \right)^{-n_2}\Bigg
  \rangle_{(1)},  \label{eq:stopA}
    \end{split}\\
    \begin{split}
  I^{q\bar{q},3}_{n_1n_2}&= \left\langle \left(p\cdot k_1 - \frac{\tau z_1}{2 (z_1-z_2)}\right)^{-n_1} \left( k_1 \cdot k_2 \right)^{-n_2}\right\rangle_{(1)} ,
    \end{split}
\end{align}
%I 
where for a given integrand $f$ we write 
\begin{align}
    \left\langle f \right\rangle_{(1)} = \int \dd[]\mathrm{PS}_S^{(2)}\  \delta \left ( \tau - 2 p \cdot k_1 - 2p\cdot k_2 \right) \
\delta (2 p\cdot k_1 - z_1 2 \bar p \cdot k_1 ) \ \delta ( 2 p\cdot k_2 - z_2 2 \bar p \cdot k_2  ) \; f.
\end{align}
We perform the IBP reduction using FIRE~\cite{fire} and obtain the following master integrals
\begin{align}
    &I^{q\bar{q},3}_{00}=\left\langle 1 \right\rangle_{(1)}, \label{eq:confAmas}
    &I^{q\bar{q},3}_{10}&=\left\langle 
    \left ( p \cdot k_1 - \frac{\tau z_1 }{2 (z_1 - z_2)} \right )^{-1}\right\rangle_{(1)}, \notag \\
    &I^{q\bar{q},3}_{01}=\left\langle
        \left(k_1 \cdot k_2\right)^{-1}\right\rangle_{(1)},
    &I^{q\bar{q},3}_{11}&=\left\langle
    \left ( p \cdot k_1 - \frac{\tau z_1 }{2 (z_1 - z_2)} \right )^{-1} \left(k_1 \cdot k_2\right)^{-1}\right\rangle_{(1)}.
  \end{align}
% w
For configuration $B$, we obtain two integral families
\begin{align}
    \begin{split}
    I^{q\bar{q},2}_{n_1n_2}&= \left\langle \left( p\cdot k_1 + \frac{\tau z_1}{2 (1-z_1)} \right)^{-n_1} \left(k_1 \cdot k_2 \right)^{-n_2}\right\rangle_{(2)}, \label{eq:stop1}
    \end{split} \\
    \begin{split}
    I^{q\bar{q},4}_{n_1n_2}&= \left\langle\left(p\cdot k_1 -\frac{\tau}{2 (1-z_2)} \right)^{-n_1} \left(k_1 \cdot k_2 \right)^{-n_2}\right\rangle_{(2)}, \label{eq:stop2}
    \end{split}
\end{align}
that are mapped on the following master integrals
\begin{align}
    &I^{q\bar{q},2}_{00}=\left\langle1\right\rangle_{(2)},
    &I^{q\bar{q},2}_{10}&=\left\langle \notag
    \left ( p\cdot k_1 +  \frac{\tau z_1 }{2 (1 - z_1)} \right )^{-1}\right\rangle_{(2)},\\
    &I^{q\bar{q},2}_{01}=\left\langle \left(k_1 \cdot k_2 \right)^{-1}\right\rangle_{(2)},
    &I^{q\bar{q},2}_{11}&=\left\langle
    \left ( p \cdot k_1 +  \frac{\tau z_1 }{2 (1 - z_1)} \right )^{-1} \left(k_1 \cdot k_2\right)^{-1}\right\rangle_{(2)}, \label{eq:confBmas} \\
    &I^{q\bar{q},4}_{10}=\left\langle
    \left ( p\cdot k_1  - \frac{\tau }{2 (1 - z_2)} \right )^{-1}\right\rangle_{(2)},
    &I^{q\bar{q}.4}_{11}&=\left\langle 
    \left ( p \cdot k_1  - \frac{\tau }{2 (1 - z_2)} \right )^{-1} \left(k_1\cdot k_2\right)^{-1}\right\rangle_{(2)}.\notag
\end{align}
In \cref{eq:stop1,eq:stop2,eq:confBmas} we used
\begin{align}
    \left\langle f \right\rangle_{(2)} =\int\dd[]\mathrm{PS}_S^{(2)} \ \delta(\tau - 2 \bar p\cdot k_1 - 2 p\cdot k_2) \
     \delta ( 2 \bar p \cdot k_1 - z_1 2 p \cdot k_1) \  \delta ( 2p\cdot k_2 - z_2 2 \bar p \cdot k_2 ) \; f.
\end{align}
We describe the calculation of the master integrals in the next section.
\subsection{Master integrals}
\label{sec:softmasters}
The master integrals shown in \cref{eq:confAmas,eq:confBmas} can be evaluated directly. When describing this calculation below, we will always assume that
$z_1 > z_2$ since all contributions to the soft function are symmetric with respect to $z_1 \leftrightarrow z_2$ permutation. 

To illustrate the simplicity of the computation, we discuss the calculation of the most complicated master integral. We provide explicit solutions to all other master integrals in \cref{sec:AppA}.
We consider the master integral
\begin{align}
    % I
    \begin{split}
    I^{q\bar{q},3}_{11} &=\int\dd[]\mathrm{PS}_S^{(2)}\  \delta \left ( \tau - 2 p \cdot k_1 - 2p\cdot k_2 \right) \
\delta (2 p\cdot k_1 - z_1 2 \bar p \cdot k_1 ) \ \delta ( 2  p\cdot k_2 - z_2 2 \bar p \cdot k_2  ) \\
&\eqspace \times \left ( p \cdot k_1 - \frac{\tau z_1 }{2 (z_1 - z_2)} \right )^{-1} \left(k_1 \cdot k_2\right)^{-1}, \label{eq:softmaster}
    \end{split}
\end{align}
The computation proceeds as follows. We begin by performing the Sudakov decomposition of the two light-like momenta $k_{1,2}$
\begin{align}
    k_{1,2} = \alpha_{1,2} \, p + \beta_{1,2} \, \bar p + k_{1,2\perp}.    
\end{align}
The integration measure $\dd[]\mathrm{PS}_S^{(2)}$ is then written as
\begin{align}
    \dd[]\mathrm{PS}_S^{(2)} =\frac{\left[\Omega^{(d-2)}\right]^{-2} }{4} \prod_{i=1}^{2}  
   {\rm d} \alpha_i\  {\rm d} \beta_i\ [ \alpha_i \beta_i ]^{-\epsilon} {\rm d}
   \Omega_i^{(d-2)}.
\end{align}
Note that integrations over $\alpha$ and $\beta$ extend from zero to infinity with constraints imposed by
$\delta$-functions.
We write
\begin{align}
    \begin{split}
I^{q \bar q, 3}_{11} &=
\frac{\left[\Omega^{(d-2)}\right]^{-2} }{4} 
2 \int  \prod_{i=1}^{2} {\rm d} \alpha_i \ {\rm d} \beta_i \ \left [ \alpha_i \beta_i \right ]^{-\epsilon}
      {\rm d} \Omega_{i}^{(d-2)} \, \delta ( \tau - \beta_1 - \beta_2) 
      \delta(\beta_1 - z_1 \alpha_1)\\
&\eqspace\times \delta ( \beta_2 - z_2 \alpha_2)\left (
            \frac{\beta_1}{2} - \frac{\tau z_1}{ 2 (z_1 - z_2)} 
     \right )^{-1}
           \left(2 k_1 \cdot k_2\right)^{-1}. \label{eq:subsin2this}
    \end{split}
\end{align}
The angular integrations in \cref{eq:subsin2this} were discussed in Ref.\cite{Monni:2011gb}. The result reads  
\begin{align}
    \begin{split}
\int \frac{{\rm d} \Omega_1^{(d-2)} {\rm d} \Omega_2^{(d-2)} }{2 k_1 \cdot k_2 }
= 
\frac{\left [ \Omega^{(d-2)} \right ]^2 }{(\sqrt{\alpha_1 \beta_2} + \sqrt{\alpha_2 \beta_1} )^2}
\ _2F_{1} \left ( 1,\frac{1}{2}-\ep,1-2\ep,\frac{4 \sqrt{\alpha_1 \alpha_2 \beta_1 \beta_2} }{
(\sqrt{\alpha_1 \beta_2} + \sqrt{\alpha_2 \beta_1} )^2
}\right ).
\label{eq16}
    \end{split}
\end{align}
Because of the delta functions in \cref{eq:subsin2this}, we need \cref{eq16} for
$\alpha_i = \beta_i/z_i$. It becomes
\begin{align}
     \begin{split}
   \int \frac{{\rm d} \Omega_1^{(d-2)} {\rm d} \Omega_2^{(d-2)} }{2 k_1 \cdot k_2 }
  \left|_{\alpha_i \to \frac{\beta_i}{z_i}} \right.
&=
\frac{\left [ \Omega^{(d-2)} \right ]^2 z_1 z_2 }{ \beta_1 \beta_2
  (\sqrt{z_1} + \sqrt{z_2} )^2}
\ _2F_{1} \left ( 1,\frac{1}{2}-\ep,1-2\ep,\frac{4 \sqrt{z_1 z_2}  }{
(\sqrt{z_1}  + \sqrt{z_2}  )^2
}\right ) 
\\
& =
\frac{\left [ \Omega^{(d-2)} \right ]^2 z_2 }{ \beta_1 \beta_2
  \left (1 + \sqrt{\frac{z_2}{z_1} }  \right )^2}
\ _2F_{1} \left ( 1,\frac{1}{2}-\ep,1-2\ep,\frac{4 \sqrt{z_2/z_1}  }{
(1+ \sqrt{z_2/z_1}  )^2 \label{eq:hgf}
}\right ).
\end{split}
\end{align}
The hypergeometric function can be simplified using the following identity
\begin{align}
\ _2F_{1} \left ( 1,\frac{1}{2}-\ep,1-2\ep,\frac{4 z  }{
(1+ z )^2
}\right ) = (1+z)^2 \ _2F_{1}\left ( 1, 1+ \ep, 1-\ep, z^2 \right ), \label{eq:shgf}
\end{align}
which is valid for $|z| < 1$.  Since we work in the region where $z_2 < z_1$, we can immediately use \cref{eq:shgf} to simplify \cref{eq:hgf}.
We obtain
\be
\begin{split} 
  & \int \frac{{\rm d} \Omega_1^{(d-2)} {\rm d} \Omega_2^{(d-2)} }{2 k_1 \cdot k_2 }
  \left|_{\alpha_i \to \frac{\beta_i}{z_i}} \right.
=
\frac{\left [ \Omega^{(d-2)} \right ]^2 z_2 }{ \beta_1 \beta_2}
\ _2F_{1} \left (1,1+\ep,1-\ep,\frac{z_2}{z_1} \right ). \label{eq:angintsoft}
\end{split}
\ee
Remarkably, the hypergeometric function in \cref{eq:angintsoft} is independent of the parameters $\alpha_i$ and $\beta_i$, allowing for
a straightforward integration. We substitute \cref{eq:angintsoft} back into \cref{eq:subsin2this}, integrate over $\alpha_1$, $\alpha_2$, $\beta_2$ and change the integration variable $\beta_1 \to \beta_1^\prime =\beta_1/\tau$. We find
\begin{align}
I^{q \bar q, 3}_{11} & =
-\frac{ \tau^{-2-4\ep}}{(z_1 z_2)^{1-\ep}}
z_2 \ _2F_{1}\left ( 1,1+\ep,1-\ep, \frac{z_2}{z_1} \right )
    \frac{(z_1 - z_2)}{z_1}
\int \limits_{0}^{1} {\rm d} \beta_1^\prime \frac{ \left ( \beta^\prime_1 (1-\beta^\prime_1) \right )^{-2\ep - 1} }
     {1 - \frac{z_1-z_2}{z_1}\beta_1^\prime }. 
\end{align}
     Upon integrating over $\beta^\prime_1$, we obtain the following result for the most complicated of the nine master integrals needed to describe the NNLO soft function
    \be
\begin{split} 
I^{q \bar q, 3}_{11} & =
-\frac{\tau^{-2-4\ep}}{(z_1 z_2)^{1-\ep}}
\frac{\Gamma^2(-2\ep)}{\Gamma(-4\ep) }  z_2 \frac{z_1 - z_2 }{z_1 }\\
& \eqspace \times \  _2F_{1}\left ( 1,1+\ep,1-\ep, \frac{z_2}{z_1} \right ) 
 \ _2F_{1} \left ( 1, -2\ep, -4 \ep, \frac{z_1-z_2}{z_1} \right ).
     \end{split} 
\ee

A complete list of master integrals can be found in \cref{sec:AppA}.
We note that for the gluon emission contribution $S^{(2)}_{gg}$ no further master integrals are required.

This concludes our discussion of the evaluation of master integrals. We discuss the remaining integration over the auxillary parameters $z_{1,2}$ in the next section.
\subsection{Integration over auxillary parameters}
\label{sec:zint}
We express the double-real contributions in terms of master integrals and write $S^{(2)}_{q\bar{q}, A}$ as follows
\begin{align}
    \begin{split}
    S^{(2)}_{q\bar{q}, A}&=2\int_0^1 \dd[]z_1 \int_0^{z_1} \dd[]z_2 \bigg[ \frac{32  \epsilon (2  \epsilon-1) z_1  z_2
    }{\tau ^2 ( z_1- z_2)^2}\; I^{q\bar{q},3}_{00}+\frac{8  \epsilon (2  \epsilon-1) z_1  z_2 ( z_1+ z_2)
    }{\tau  ( z_1- z_2)^3}\; I^{q\bar{q},3}_{10}  \\
   &\eqspace+\frac{8 ( z_1+ z_2) \left(16  \epsilon^3 z_1  z_2- \epsilon^2
    ( z_1+ z_2)^2+ \epsilon \left( z_1^2-6 z_1  z_2+ z_2^2\right)+z_1  z_2\right)
    }{(4  \epsilon-1) ( z_1- z_2)^4}\; I^{q\bar{q},3}_{01}  \\ 
    &\eqspace+\frac{8 \tau  z_1  z_2 \left( \epsilon^2 ( z_1+ z_2)^2-z_1  z_2\right)
    }{( z_1- z_2)^5}\; I^{q\bar{q},3}_{11} \bigg]. \label{eq:redcofAp}
    \end{split}
\end{align}
It appears that upon 
substituting solutions for the master integrals \crefrange{eq:ma2}{eq:ma1} into \cref{eq:redcofAp}, we will have to perform non-trivial integrations
over $z_1$ and $z_2$. However, after changing variables $z_2=t \ z_1$, the $z_1$ integration factors out. The remaining $t$ integration seems to include terms that are proportional
to $(1-t)^{-4}$. However, upon taking the limit $t \to 1$ we find that the most singular term actually scales like $(1-t)^{-1-2\epsilon}$ and, therefore, can be easily subtracted.
We perform an endpoint subtraction at $t=1$, expand the integrand in a Laurent series in $\epsilon$ and compute the integral order by order in $\epsilon$ with
the help of HyperInt~\cite{Panzer:2014caa}. The final result reads
\begin{align}
    \begin{split}
    S^{(2)}_{q\bar{q}, A}&=\tau^{-1-4\epsilon} \Bigg[-\frac{2}{3 \epsilon^2}-\frac{10}{9 \epsilon}-\frac{38}{27}-\frac{2 \pi ^2}{9}+\epsilon \left(-\frac{16 \zeta (3)}{3}-\frac{238}{81}-\frac{10 \pi ^2}{27}\right)\\
    &\eqspace+\epsilon^2 \left(-\frac{80 \zeta (3)}{9}-\frac{962}{243}-\frac{92 \pi ^2}{81}-\frac{8 \pi ^4}{45}\right)  \\ 
    &\eqspace+\epsilon^3 \left(-\frac{736 \zeta (3)}{27}+\frac{104 \pi ^2 \zeta (3)}{9}-168 \zeta (5)+\frac{4394}{729}-\frac{832 \pi ^2}{243}-\frac{8 \pi
    ^4}{27}\right)+\order{\epsilon^4}  \Bigg] .
    \end{split}
\end{align}
The physical meaning of the auxillary variables $z_1$ and $z_2$ can be understood by considering the Sudakov decomposition of $k_{1,2}$. The singularity at $z_1=z_2$ describes the
limit were the quark and the anti-quark become collinear to each other, while the $z_1=0$, singularity describes the kinematic configuration in which the gluon, that emits the $q\bar{q}$ pair, becomes collinear to the
light-like directions $p^\mu$. The $\tau  \to 0$ limit controls the double-soft divergence. 

Next, we discuss the contribution $S^{(2)}_{q\bar{q}, B}$ that describes the emission of a $q\bar{q}$ pair in configuration $B$. Written in terms of master integrals, this contribution reads
\begin{align}
    S^{(2)}_{q\bar{q}, B}&=2\int_0^1 \dd[]z_1 \int_0^{z_1} \dd[]z_2 \Bigg[\frac{32 \epsilon (4 \epsilon-1) z_1 z_2}{\tau ^2 (z_1 z_2-1)^2}\; I^{q\bar{q},2}_{00} \notag 
    \\ &\eqspace -\frac{8 \tau  z_2 \left(\epsilon^2 (z_2+1) \left(z_1^2 z_2^2-1\right)+\epsilon (z_1-1) z_2 (z_1 z_2+1)-z_1 (z_2-1) z_2\right)
   }{(z_2-1)^3 (z_1 z_2-1)^3}\;I^{q\bar{q},4}_{11} \notag \\ &\eqspace+  \Bigg(\frac{16 z_1 z_2 (z_1 (-z_2)+z_1+z_2-1)}{(z_1-1)^2 (z_2-1)^2 (z_1 z_2-1)^2} \notag \\
   &\eqspace+\epsilon \frac{8 \left(z_1^3 (-(z_2-3)) z_2^2+z_1^2 z_2 \left(3 z_2^2-11 z_2+6\right)+z_1 \left(6 z_2^2-11 z_2+3\right)+3 z_2-1\right)}{(z_1-1)^2
   (z_2-1)^2 (z_1 z_2-1)^2} \notag \\
   &\eqspace+\epsilon^2 \frac{32 \left(z_1^3 z_2^2+z_1^2 z_2 \left(z_2^2-4 z_2+2\right)+z_1 \left(2 z_2^2-4 z_2+1\right)+z_2\right)}{(z_1-1)^2 (z_2-1)^2 (z_1
   z_2-1)^2} \Bigg)\; I^{q\bar{q},2}_{01} \label{eq:softBfull} \\ &\eqspace+\frac{8 \tau  z_1 \left(\epsilon
    z_1^2 z_2^2 (\epsilon z_1+\epsilon+1)-(\epsilon+1) (z_1-1) z_1 z_2-\epsilon (\epsilon z_1+\epsilon+z_1)\right)}{(z_1-1)^3
    (z_1 z_2-1)^3}\;I^{q\bar{q},2}_{11} \notag \\ &\eqspace+\frac{8 \epsilon z_1 z_2 (2 \epsilon (z_1+1) (z_1 z_2-1)-(z_1-1) (z_1 z_2+1))}{\tau  (z_1-1)
    (z_1 z_2-1)^3}\;I^{q\bar{q},2}_{10} \notag \\ &\eqspace-\frac{8 \epsilon z_1 z_2 (2 \epsilon (z_2+1) (z_1 z_2-1)-(z_2-1) (z_1 z_2+1))}{\tau  (z_2-1)
    (z_1 z_2-1)^3}\;I^{q\bar{q},4}_{10} \Bigg] \notag.
\end{align}
While the expression in \cref{eq:softBfull} appears to be even more complicated than the one in \cref{eq:redcofAp}, it is actually much simpler.
This can be expected since, in configuration $B$, the quark and the anti-quark are emitted into different hemispheres. Thus both $z_1=0$ and $z_1=z_2$ (collinear) singularities should be absent.
We therefore expect that we can simply expand the integrand in \cref{eq:softBfull} in a Laurent series in $\epsilon$ and integrate the result order by order in that expansion. This is indeed what happens.
 The final result reads
\begin{align}
    \begin{split}
S^{(2)}_{q\bar{q}, B}&=\tau^{-1-4\epsilon} \Bigg[
    \frac{4 \pi ^2}{9}-\frac{2}{3}+\epsilon \left(\frac{56 \zeta (3)}{3}+\frac{22}{9}-\frac{32 \pi ^2}{27}\right)  
    \\ &\eqspace +\epsilon^2 \left(-\frac{448 \zeta (3)}{9}+\frac{226}{27}+\frac{172 \pi ^2}{81}+\frac{34 \pi ^4}{45}\right)\\ &\eqspace +\epsilon^3 \left(\frac{2480 \zeta (3)}{27}-\frac{88 \pi ^2 \zeta (3)}{3}+\frac{1640 \zeta (5)}{3}+\frac{1438}{81}-\frac{668 \pi
    ^2}{243}-\frac{272 \pi ^4}{135}\right)+\order{\epsilon^4} \Bigg].
    \end{split}
\end{align}

The calculation of $S_{gg}^{(2)}$ can be performed in the same way. While the gluon emission amplitudes include an additional singular 
configuration compared to the $q\bar{q}$ case, the $z_{1,2}$ singularity structure remains unchanged.
The additional ``single-soft'' divergence, which is absent in $q\bar{q}$ emission,
is accounted for by an additional factor $\epsilon^{-1}$  that originates from the IBP reduction, and thus the complexity of the $z_{1,2}$ integrations remains unchanged.
% R
We present our results for the soft function in the next section.
\subsection{Results}
\label{sec:resultssoft}
We now present our final result for the bare soft function $S^{(2)}$ through $\order{\epsilon^2}$ at NNLO QCD.
To this end, we write
 \begin{align}
     S^{(2)}&=\tau^{-1-4 \epsilon}\Bigg( C_a^2 \ S^{(2)}_A + C_a T_Fn_f \ S^{(2)}_B
     +C_A C_a \ S^{(2)}_C\Bigg) \label{eq:soft2final}.
 \end{align} 
The individual contributions shown in \cref{eq:soft2final} read
 \begin{align}
    \begin{split}
    S^{(2)}_A&= -\frac{8}{\epsilon^3}+\frac{16 \pi ^2}{3 \epsilon}+128 \zeta (3)+ \epsilon \frac{16 \pi ^4}{5}+\epsilon^2 \left(1536 \zeta (5)-\frac{256 \pi ^2 \zeta (3)}{3}\right) \\
    &\eqspace+ \epsilon^3 \left( \frac{2528 \pi ^6}{945}-1024 \zeta (3)^2 \right), \label{eq:softfinala}
    \end{split}\\
    \begin{split}
    S^{(2)}_B&=-\frac{4}{3 \epsilon^2}-\frac{20}{9 \epsilon}+\frac{4 \pi ^2}{9}-\frac{112}{27}+\epsilon \left(\frac{80 \zeta (3)}{3}-\frac{80}{81}-\frac{28 \pi ^2}{9}\right)  \\
    &\eqspace+ \epsilon^2 \left(-\frac{352 \zeta (3)}{3}+\frac{2144}{243}+\frac{160 \pi ^2}{81}+\frac{52 \pi ^4}{45} \right)  \\
    &\eqspace+ \epsilon^3 \left(\frac{3488 \zeta (3)}{27}-\frac{320 \pi ^2 \zeta (3)}{9}+\frac{2272 \zeta (5)}{3}+\frac{34672}{729}-\frac{1000 \pi
    ^2}{81}-\frac{208 \pi ^4}{45}\right),\label{eq:softfinalb}
    \end{split}\\
    \begin{split}
    S^{(2)}_C&=\frac{11}{3 \epsilon^2}+\frac{1}{\epsilon}\left(\frac{67}{9}-\frac{\pi ^2}{3} \right)-14 \zeta (3)+\frac{404}{27}-\frac{11 \pi ^2}{9}\\
    &\eqspace+\epsilon\left(-\frac{220 \zeta (3)}{3}+\frac{2140}{81}+\frac{67 \pi ^2}{9}-\frac{49 \pi ^4}{90}\right)  \\ 
    &\eqspace+\epsilon^2 \left( 268 \zeta (3)+\frac{8 \pi ^2 \zeta (3)}{3}-170 \zeta (5)+\frac{12416}{243}-\frac{368 \pi ^2}{81}-\frac{143 \pi ^4}{45}\right)  \\
    &\eqspace+\epsilon^3 \Biggl(-\frac{7864 \zeta (3)}{27}+\frac{880 \pi ^2 \zeta (3)}{9}-126 \zeta (3)^2-\frac{6248 \zeta (5)}{3}\\
    &\eqspace \eqspace+\frac{67528}{729}+\frac{2416
    \pi ^2}{81}+\frac{469 \pi ^4}{45}-\frac{10 \pi ^6}{63} \Biggr).\label{eq:softfinalc}
          \end{split}
 \end{align}
 We set $C_a=C_F$, compare the result \crefrange{eq:soft2final}{eq:softfinalc} against the $\order{\epsilon^0}$ results in Refs.~\cite{Monni:2011gb,Kelley:2011ng} and find full agreement.
A computer-readable expression for the bare soft function \cref{eq:soft2final}
is contained in the ancillary file provided with this submission. 

% -------------------------

% -------------------------
\section{Conclusion}
\label{sec:con}
We computed all NNLO zero-jettiness beam functions and the soft function
expanded through ${\cal O}(\epsilon^2)$ using soft and collinear limits of QCD amplitudes, reverse unitarity and IBP relations.
Our results provide one of the building blocks for calculating the N$^3$LO soft function and beam function matching coefficients; some results for the beam functions described here have already been used in Ref.~\cite{Behring:2019quf}.

While the N$^3$LO QCD computations of beam functions~\cite{Behring:2019quf,Luo:2019szz} are the first steps towards implementing zero-jettiness slicing to describe color-singlet production in hadron collisions, a significant amount of work remains to be done. Indeed, in addition to going beyond the large-$N_c$ approximation
other matching coefficients $I_{q_i,g}$, $I_{q_i,\bar{q}_j}$, $I_{g,g}$ and $I_{g,q_i}$ have to be calculated.
Furthermore, the N$^3$LO zero-jettiness soft function is currently unknown. Since the computation of the soft function is complicated by step functions
in the phase-space measure, it is important to understand how to connect it to modern computational methods that involve IBP reductions and differential equations.
The method discussed in this paper is a first step in that direction.

% -------------------------
\section*{Acknowledgements}
I wish to thank Kirill Melnikov for inspiring discussions on the subject of this paper, as well as invaluable comments on the manuscript.
I am grateful to Arnd Behring, Maximilian Delto, Christopher Wever and Robbert Rietkerk for stimulating discussions and help with various aspects of the calculation.
The support by the Deutsche Forschungsgemeinschaft (DFG, German Research Foundation) under grant 396021762 - TRR 257 and the Doctoral School „Karlsruhe School of Elementary and Astroparticle Physics: Science and Technology“~(KSETA) is gratefully acknowledged.
%

% -------------------------
\appendix
\section{Appendix A}
\label{sec:AppA}
In this Appendix, we present explicit intermediate results for the calculation of the soft function, that were omitted in \cref{sec:soft}.
In that section we split the double-real contributions into two configurations for the emitted partons, $A$ and $B$.
The complete set of master integrals that describe configuration $A$ read 
\begin{align}
    I^{q \bar q, 3}_{00}
 & = \frac{1}{4} \frac{\tau^{1-4\epsilon} }{(z_1 z_2)^{1-\epsilon}}
\frac{\Gamma^2(1-2\epsilon)}{ \Gamma(2-4\epsilon) }, \label{eq:ma2}\\
I^{q \bar q, 3}_{10} & =  
-\frac{1}{2} \frac{\tau^{-4\epsilon} }{(z_1 z_2)^{1-\epsilon}}
\frac{z_1 -z_2}{z_1} \frac{\Gamma^2(1-2\ep)}{\Gamma(2-4\ep)}
\ _2F_{1} \left ( 1, 1-2\epsilon, 2-4\epsilon, \frac{z_1 - z_2}{z_1} \right ), \label{eq:ma3}\\
I^{q \bar q, 3}_{01}
&= \frac{1}{2} \frac{\tau^{-1-4\epsilon} }{(z_1 z_2)^{1-\epsilon}}
\frac{\Gamma^2(-2\ep)}{\Gamma(-4\ep)} \;
z_2 F_{12}\left ( 1,1+\ep,1-\ep,\frac{z_2}{z_1} \right ), \label{eq:ma4}\\
\begin{split}
I^{q \bar q, 3}_{11} & =-\frac{ \tau^{-2-4\ep}}{(z_1 z_2)^{1-\ep}}
\frac{\Gamma^2(-2\ep)}{\Gamma(-4\ep) } z_2 \frac{z_1 - z_2 }{z_1 }\\
& \eqspace \times  \ _2F_{1}\left ( 1,1+\ep,1-\ep, \frac{z_2}{z_1} \right ) 
 \ _2F_{1} \left ( 1, -2\ep, -4 \ep, \frac{z_1-z_2}{z_1} \right ).
     \label{eq:ma1}
\end{split}
\end{align}
For the configuration $B$ the master integrals read
\begin{align}
    I^{q \bar q, 2}_{00} & =    I^{q \bar q, 3}_{00} \label{eq:ma5} \\
    I^{q \bar q, 2}_{10}  &= 
    \frac{1}{2} \frac{\tau^{-4\ep}}{ (z_1 z_2)^{1-\ep} } \frac{1-z_1}{z_1}
    \frac{\Gamma^2(1-2\ep)}{\Gamma(2-4\ep)} \ _2F_{1}\left ( 1, 1-2\ep,2-4\ep,-\frac{1-z_1}{z_1} \right ), \label{eq:ma6} \\
    I^{q \bar q,2}_{01} &= \frac{1}{2}
\frac{\tau^{-1-4\ep} z_1 z_2 }{(z_1 z_2)^{1-\ep}}  \frac{\Gamma^2(-2\ep)}{\Gamma(-4\ep)}
\ _2F_{1} \left ( 1, 1+ \ep, 1-\ep, z_1 z_2 \right ),  \label{eq:ma7}\\
\begin{split}
I_{11}^{q \bar q, 2} &=
\frac{\tau^{-2-4\ep} (1-z_1) z_2 }{ (z_1 z_2)^{1-\ep}}
\frac{\Gamma^2(-2\ep)}{\Gamma(-4\ep)}  \\
&\eqspace \times \ _2F_{1} \left ( 1, 1+ \ep, 1-\ep, z_1 z_2 \right )
\ _2F_{1} \left ( 1, -2\ep, -4\ep, -\frac{1-z_1}{z_1} \right ),  \label{eq:ma8}
\end{split}\\
I^{q \bar q, 4}_{10}
&= -\frac{1}{2} \frac{ \tau^{-4\ep}}{ (z_1 z_2)^{1-\ep} }
(1-z_2) \frac{\Gamma^2(1-2\ep) }{\Gamma(2-4\ep)} \ _2F_{1}(1,1-2\ep,2-4\ep,1-z_2),  \label{eq:ma9}\\
I^{q \bar q, 4}_{11}
 & = - \frac{\tau^{-2-4\ep} z_1 z_2 (1-z_2) \label{eq:ma10}
}{ (z_1 z_2)^{1-\ep} }  \frac{\Gamma^2(-2\ep) }{\Gamma(-4\ep)} \notag
\\
 &\eqspace \times \ _2F_{1} \left ( 1, 1+ \ep, 1-\ep, z_1 z_2 \right ) \ _2F_{1}(1,-2\ep,-4\ep,1-z_2).
\end{align}

% -------------------------
\bibliographystyle{utphys}
\bibliography{Bibliography/thesis.bib}{}

\providecommand{\href}[2]{#2}\begingroup\raggedright\begin{thebibliography}{10}

\bibitem{Tackmann2014}
J.~R. Gaunt, M.~Stahlhofen, and F.~J. Tackmann, ``{The Quark Beam Function at
  Two Loops},'' \href{http://dx.doi.org/10.1007/JHEP04(2014)113}{{\em JHEP}
  {\bfseries 04} (2014) 113},
\href{http://arxiv.org/abs/1401.5478}{{\ttfamily arXiv:1401.5478 [hep-ph]}}.
%%CITATION = ARXIV:1401.5478;%%.

\bibitem{Gaunt:2014cfa}
J.~Gaunt, M.~Stahlhofen, and F.~J. Tackmann, ``{The Gluon Beam Function at Two
  Loops},'' \href{http://dx.doi.org/10.1007/JHEP08(2014)020}{{\em JHEP}
  {\bfseries 08} (2014) 020},
\href{http://arxiv.org/abs/1405.1044}{{\ttfamily arXiv:1405.1044 [hep-ph]}}.
%%CITATION = ARXIV:1405.1044;%%.

\bibitem{Petriello}
R.~Boughezal, F.~Petriello, U.~Schubert, and H.~Xing, ``{Spin-dependent quark
  beam function at NNLO},''
  \href{http://dx.doi.org/10.1103/PhysRevD.96.034001}{{\em Phys. Rev.}
  {\bfseries D96} no.~3, (2017) 034001},
\href{http://arxiv.org/abs/1704.05457}{{\ttfamily arXiv:1704.05457 [hep-ph]}}.
%%CITATION = ARXIV:1704.05457;%%.

\bibitem{Monni:2011gb}
P.~F. Monni, T.~Gehrmann, and G.~Luisoni, ``{Two-Loop Soft Corrections and
  Resummation of the Thrust Distribution in the Dijet Region},''
  \href{http://dx.doi.org/10.1007/JHEP08(2011)010}{{\em JHEP} {\bfseries 08}
  (2011) 010},
\href{http://arxiv.org/abs/1105.4560}{{\ttfamily arXiv:1105.4560 [hep-ph]}}.
%%CITATION = ARXIV:1105.4560;%%.

\bibitem{Kelley:2011ng}
R.~Kelley, M.~D. Schwartz, R.~M. Schabinger, and H.~X. Zhu, ``{The two-loop
  hemisphere soft function},''
  \href{http://dx.doi.org/10.1103/PhysRevD.84.045022}{{\em Phys. Rev.}
  {\bfseries D84} (2011) 045022},
\href{http://arxiv.org/abs/1105.3676}{{\ttfamily arXiv:1105.3676 [hep-ph]}}.
%%CITATION = ARXIV:1105.3676;%%.

\bibitem{Catani:2007vq}
S.~Catani and M.~Grazzini, ``{An NNLO subtraction formalism in hadron
  collisions and its application to Higgs boson production at the LHC},''
  \href{http://dx.doi.org/10.1103/PhysRevLett.98.222002}{{\em Phys. Rev. Lett.}
  {\bfseries 98} (2007) 222002},
\href{http://arxiv.org/abs/hep-ph/0703012}{{\ttfamily arXiv:hep-ph/0703012
  [hep-ph]}}.
%%CITATION = HEP-PH/0703012;%%.

\bibitem{N2LO4}
R.~Boughezal, C.~Focke, X.~Liu, and F.~Petriello, ``{$W$-boson production in
  association with a jet at next-to-next-to-leading order in perturbative
  QCD},'' \href{http://dx.doi.org/10.1103/PhysRevLett.115.062002}{{\em Phys.
  Rev. Lett.} {\bfseries 115} no.~6, (2015) 062002},
\href{http://arxiv.org/abs/1504.02131}{{\ttfamily arXiv:1504.02131 [hep-ph]}}.
%%CITATION = ARXIV:1504.02131;%%.

\bibitem{Gaunt:2015pea}
J.~Gaunt, M.~Stahlhofen, F.~J. Tackmann, and J.~R. Walsh, ``{N-jettiness
  Subtractions for NNLO QCD Calculations},''
  \href{http://dx.doi.org/10.1007/JHEP09(2015)058}{{\em JHEP} {\bfseries 09}
  (2015) 058},
\href{http://arxiv.org/abs/1505.04794}{{\ttfamily arXiv:1505.04794 [hep-ph]}}.
%%CITATION = ARXIV:1505.04794;%%.

\bibitem{Bonciani:2015sha}
R.~Bonciani, S.~Catani, M.~Grazzini, H.~Sargsyan, and A.~Torre, ``{The $q_T$
  subtraction method for top quark production at hadron colliders},''
  \href{http://dx.doi.org/10.1140/epjc/s10052-015-3793-y}{{\em Eur. Phys. J.}
  {\bfseries C75} no.~12, (2015) 581},
\href{http://arxiv.org/abs/1508.03585}{{\ttfamily arXiv:1508.03585 [hep-ph]}}.
%%CITATION = ARXIV:1508.03585;%%.

\bibitem{Catani:2009sm}
S.~Catani, L.~Cieri, G.~Ferrera, D.~de~Florian, and M.~Grazzini, ``{Vector
  boson production at hadron colliders: a fully exclusive QCD calculation at
  NNLO},'' \href{http://dx.doi.org/10.1103/PhysRevLett.103.082001}{{\em Phys.
  Rev. Lett.} {\bfseries 103} (2009) 082001},
\href{http://arxiv.org/abs/0903.2120}{{\ttfamily arXiv:0903.2120 [hep-ph]}}.
%%CITATION = ARXIV:0903.2120;%%.

\bibitem{N2LO1}
R.~Boughezal, C.~Focke, W.~Giele, X.~Liu, and F.~Petriello, ``{Higgs boson
  production in association with a jet at NNLO using jettiness subtraction},''
  \href{http://dx.doi.org/10.1016/j.physletb.2015.06.055}{{\em Phys. Lett.}
  {\bfseries B748} (2015) 5--8},
\href{http://arxiv.org/abs/1505.03893}{{\ttfamily arXiv:1505.03893 [hep-ph]}}.
%%CITATION = ARXIV:1505.03893;%%.

\bibitem{Boughezal:2015ded}
R.~Boughezal, J.~M. Campbell, R.~K. Ellis, C.~Focke, W.~T. Giele, X.~Liu, and
  F.~Petriello, ``{Z-boson production in association with a jet at
  next-to-next-to-leading order in perturbative QCD},''
  \href{http://dx.doi.org/10.1103/PhysRevLett.116.152001}{{\em Phys. Rev.
  Lett.} {\bfseries 116} no.~15, (2016) 152001},
\href{http://arxiv.org/abs/1512.01291}{{\ttfamily arXiv:1512.01291 [hep-ph]}}.
%%CITATION = ARXIV:1512.01291;%%.

\bibitem{Boughezal:2016wmq}
R.~Boughezal, J.~M. Campbell, R.~K. Ellis, C.~Focke, W.~Giele, X.~Liu,
  F.~Petriello, and C.~Williams, ``{Color singlet production at NNLO in
  MCFM},'' \href{http://dx.doi.org/10.1140/epjc/s10052-016-4558-y}{{\em Eur.
  Phys. J.} {\bfseries C77} no.~1, (2017) 7},
\href{http://arxiv.org/abs/1605.08011}{{\ttfamily arXiv:1605.08011 [hep-ph]}}.
%%CITATION = ARXIV:1605.08011;%%.

\bibitem{Boughezal:2016dtm}
R.~Boughezal, X.~Liu, and F.~Petriello, ``{W-boson plus jet differential
  distributions at NNLO in QCD},''
  \href{http://dx.doi.org/10.1103/PhysRevD.94.113009}{{\em Phys. Rev.}
  {\bfseries D94} no.~11, (2016) 113009},
\href{http://arxiv.org/abs/1602.06965}{{\ttfamily arXiv:1602.06965 [hep-ph]}}.
%%CITATION = ARXIV:1602.06965;%%.

\bibitem{Grazzini:2016ctr}
M.~Grazzini, S.~Kallweit, S.~Pozzorini, D.~Rathlev, and M.~Wiesemann,
  ``{$W^{+}W^{-}$ production at the LHC: fiducial cross sections and
  distributions in NNLO QCD},''
  \href{http://dx.doi.org/10.1007/JHEP08(2016)140}{{\em JHEP} {\bfseries 08}
  (2016) 140},
\href{http://arxiv.org/abs/1605.02716}{{\ttfamily arXiv:1605.02716 [hep-ph]}}.
%%CITATION = ARXIV:1605.02716;%%.

\bibitem{Grazzini:2016swo}
M.~Grazzini, S.~Kallweit, D.~Rathlev, and M.~Wiesemann, ``{$W^{\pm}Z$
  production at hadron colliders in NNLO QCD},''
  \href{http://dx.doi.org/10.1016/j.physletb.2016.08.017}{{\em Phys. Lett.}
  {\bfseries B761} (2016) 179--183},
\href{http://arxiv.org/abs/1604.08576}{{\ttfamily arXiv:1604.08576 [hep-ph]}}.
%%CITATION = ARXIV:1604.08576;%%.

\bibitem{Grazzini:2017mhc}
M.~Grazzini, S.~Kallweit, and M.~Wiesemann, ``{Fully differential NNLO
  computations with MATRIX},''
  \href{http://dx.doi.org/10.1140/epjc/s10052-018-5771-7}{{\em Eur. Phys. J.}
  {\bfseries C78} no.~7, (2018) 537},
\href{http://arxiv.org/abs/1711.06631}{{\ttfamily arXiv:1711.06631 [hep-ph]}}.
%%CITATION = ARXIV:1711.06631;%%.

\bibitem{Grazzini:2017ckn}
M.~Grazzini, S.~Kallweit, D.~Rathlev, and M.~Wiesemann, ``{$W^\pm Z$ production
  at the LHC: fiducial cross sections and distributions in NNLO QCD},''
  \href{http://dx.doi.org/10.1007/JHEP05(2017)139}{{\em JHEP} {\bfseries 05}
  (2017) 139},
\href{http://arxiv.org/abs/1703.09065}{{\ttfamily arXiv:1703.09065 [hep-ph]}}.
%%CITATION = ARXIV:1703.09065;%%.

\bibitem{Catani:2018krb}
S.~Catani, L.~Cieri, D.~de~Florian, G.~Ferrera, and M.~Grazzini, ``{Diphoton
  production at the LHC: a QCD study up to NNLO},''
  \href{http://dx.doi.org/10.1007/JHEP04(2018)142}{{\em JHEP} {\bfseries 04}
  (2018) 142},
\href{http://arxiv.org/abs/1802.02095}{{\ttfamily arXiv:1802.02095 [hep-ph]}}.
%%CITATION = ARXIV:1802.02095;%%.

\bibitem{Catani:2019hip}
S.~Catani, S.~Devoto, M.~Grazzini, S.~Kallweit, and J.~Mazzitelli, ``{Top-quark
  pair production at the LHC: Fully differential QCD predictions at NNLO},''
  \href{http://dx.doi.org/10.1007/JHEP07(2019)100}{{\em JHEP} {\bfseries 07}
  (2019) 100},
\href{http://arxiv.org/abs/1906.06535}{{\ttfamily arXiv:1906.06535 [hep-ph]}}.
%%CITATION = ARXIV:1906.06535;%%.

\bibitem{Boughezal:2019ggi}
R.~Boughezal, A.~Isgrò, and F.~Petriello, ``{Next-to-leading power corrections
  to $V+1$ jet production in $N$-jettiness subtraction},''
  \href{http://dx.doi.org/10.1103/PhysRevD.101.016005}{{\em Phys. Rev.}
  {\bfseries D101} no.~1, (2020) 016005},
\href{http://arxiv.org/abs/1907.12213}{{\ttfamily arXiv:1907.12213 [hep-ph]}}.
%%CITATION = ARXIV:1907.12213;%%.

\bibitem{Stewart:2010tn}
I.~W. Stewart, F.~J. Tackmann, and W.~J. Waalewijn, ``{N-Jettiness: An
  Inclusive Event Shape to Veto Jets},''
  \href{http://dx.doi.org/10.1103/PhysRevLett.105.092002}{{\em Phys. Rev.
  Lett.} {\bfseries 105} (2010) 092002},
\href{http://arxiv.org/abs/1004.2489}{{\ttfamily arXiv:1004.2489 [hep-ph]}}.
%%CITATION = ARXIV:1004.2489;%%.

\bibitem{Stewart:2009yx}
I.~W. Stewart, F.~J. Tackmann, and W.~J. Waalewijn, ``{Factorization at the
  LHC: From PDFs to Initial State Jets},''
  \href{http://dx.doi.org/10.1103/PhysRevD.81.094035}{{\em Phys. Rev.}
  {\bfseries D81} (2010) 094035},
\href{http://arxiv.org/abs/0910.0467}{{\ttfamily arXiv:0910.0467 [hep-ph]}}.
%%CITATION = ARXIV:0910.0467;%%.

\bibitem{Baikov:2009bg}
P.~A. Baikov, K.~G. Chetyrkin, A.~V. Smirnov, V.~A. Smirnov, and
  M.~Steinhauser, ``{Quark and gluon form factors to three loops},''
  \href{http://dx.doi.org/10.1103/PhysRevLett.102.212002}{{\em Phys. Rev.
  Lett.} {\bfseries 102} (2009) 212002},
\href{http://arxiv.org/abs/0902.3519}{{\ttfamily arXiv:0902.3519 [hep-ph]}}.
%%CITATION = ARXIV:0902.3519;%%.

\bibitem{Gehrmann:2010ue}
T.~Gehrmann, E.~W.~N. Glover, T.~Huber, N.~Ikizlerli, and C.~Studerus,
  ``{Calculation of the quark and gluon form factors to three loops in QCD},''
  \href{http://dx.doi.org/10.1007/JHEP06(2010)094}{{\em JHEP} {\bfseries 06}
  (2010) 094},
\href{http://arxiv.org/abs/1004.3653}{{\ttfamily arXiv:1004.3653 [hep-ph]}}.
%%CITATION = ARXIV:1004.3653;%%.

\bibitem{Behring:2019quf}
A.~Behring, K.~Melnikov, R.~Rietkerk, L.~Tancredi, and C.~Wever, ``{Quark beam
  function at next-to-next-to-next-to-leading order in perturbative QCD in the
  generalized large-$N_c$ approximation},''
  \href{http://dx.doi.org/10.1103/PhysRevD.100.114034}{{\em Phys. Rev.}
  {\bfseries D100} no.~11, (2019) 114034},
\href{http://arxiv.org/abs/1910.10059}{{\ttfamily arXiv:1910.10059 [hep-ph]}}.
%%CITATION = ARXIV:1910.10059;%%.

\bibitem{Luo:2019szz}
M.-x. Luo, T.-Z. Yang, H.~X. Zhu, and Y.~J. Zhu, ``{Quark Transverse Parton
  Distribution at the Next-to-Next-to-Next-to-Leading Order},''
\href{http://arxiv.org/abs/1912.05778}{{\ttfamily arXiv:1912.05778 [hep-ph]}}.
%%CITATION = ARXIV:1912.05778;%%.

\bibitem{Baranowski}
D.~Baranowski, ``{Quark beam function at NNLO to higher orders in epsilon},''
  {\em Master's thesis, KIT} (2019) .

\bibitem{Catani:1999ss}
S.~Catani and M.~Grazzini, ``{Infrared factorization of tree level QCD
  amplitudes at the next-to-next-to-leading order and beyond},''
  \href{http://dx.doi.org/10.1016/S0550-3213(99)00778-6}{{\em Nucl. Phys.}
  {\bfseries B570} (2000) 287--325},
\href{http://arxiv.org/abs/hep-ph/9908523}{{\ttfamily arXiv:hep-ph/9908523
  [hep-ph]}}.
%%CITATION = HEP-PH/9908523;%%.

\bibitem{Catani:2000pi}
S.~Catani and M.~Grazzini, ``{The soft gluon current at one loop order},''
  \href{http://dx.doi.org/10.1016/S0550-3213(00)00572-1}{{\em Nucl. Phys.}
  {\bfseries B591} (2000) 435--454},
\href{http://arxiv.org/abs/hep-ph/0007142}{{\ttfamily arXiv:hep-ph/0007142
  [hep-ph]}}.
%%CITATION = HEP-PH/0007142;%%.

\bibitem{reverseunit}
C.~Anastasiou and K.~Melnikov, ``{Higgs boson production at hadron colliders in
  NNLO QCD},'' \href{http://dx.doi.org/10.1016/S0550-3213(02)00837-4}{{\em
  Nucl. Phys.} {\bfseries B646} (2002) 220--256},
\href{http://arxiv.org/abs/hep-ph/0207004}{{\ttfamily arXiv:hep-ph/0207004
  [hep-ph]}}.
%%CITATION = HEP-PH/0207004;%%.

\bibitem{ibp}
K.~G. Chetyrkin and F.~V. Tkachov, ``{Integration by Parts: The Algorithm to
  Calculate beta Functions in 4 Loops},''
\href{http://dx.doi.org/10.1016/0550-3213(81)90199-1}{{\em Nucl. Phys.}
  {\bfseries B192} (1981) 159--204}.
%%CITATION = NUPHA,B192,159;%%.

\bibitem{Waalewijn}
M.~Ritzmann and W.~J. Waalewijn, ``{Fragmentation in Jets at NNLO},''
  \href{http://dx.doi.org/10.1103/PhysRevD.90.054029}{{\em Phys. Rev.}
  {\bfseries D90} no.~5, (2014) 054029},
\href{http://arxiv.org/abs/1407.3272}{{\ttfamily arXiv:1407.3272 [hep-ph]}}.
%%CITATION = ARXIV:1407.3272;%%.

\bibitem{Harlander:2020cyh}
R.~V. Harlander, S.~Y. Klein, and M.~Lipp, ``{FeynGame},''
\href{http://arxiv.org/abs/2003.00896}{{\ttfamily arXiv:2003.00896
  [physics.ed-ph]}}.
%%CITATION = ARXIV:2003.00896;%%.

\bibitem{fire}
A.~V. Smirnov and F.~S. Chuharev, ``{FIRE6: Feynman Integral REduction with
  Modular Arithmetic},''
\href{http://arxiv.org/abs/1901.07808}{{\ttfamily arXiv:1901.07808 [hep-ph]}}.
%%CITATION = ARXIV:1901.07808;%%.

\bibitem{vanNeerven:1985xr}
W.~L. van Neerven, ``{Dimensional Regularization of Mass and Infrared
  Singularities in Two Loop On-shell Vertex Functions},''
\href{http://dx.doi.org/10.1016/0550-3213(86)90165-3}{{\em Nucl. Phys.}
  {\bfseries B268} (1986) 453--488}.
%%CITATION = NUPHA,B268,453;%%.

\bibitem{Somogyi:2011ir}
G.~Somogyi, ``{Angular integrals in d dimensions},''
  \href{http://dx.doi.org/10.1063/1.3615515}{{\em J. Math. Phys.} {\bfseries
  52} (2011) 083501},
\href{http://arxiv.org/abs/1101.3557}{{\ttfamily arXiv:1101.3557 [hep-ph]}}.
%%CITATION = ARXIV:1101.3557;%%.

\bibitem{abramowitz+stegun}
M.~Abramowitz and I.~A. Stegun, {\em Handbook of Mathematical Functions with
  Formulas, Graphs, and Mathematical Tables}.
\newblock Dover, New York, 1964.

\bibitem{Huber:2005yg}
T.~Huber and D.~Maitre, ``{HypExp: A Mathematica package for expanding
  hypergeometric functions around integer-valued parameters},''
  \href{http://dx.doi.org/10.1016/j.cpc.2006.01.007}{{\em Comput. Phys.
  Commun.} {\bfseries 175} (2006) 122--144},
\href{http://arxiv.org/abs/hep-ph/0507094}{{\ttfamily arXiv:hep-ph/0507094
  [hep-ph]}}.
%%CITATION = HEP-PH/0507094;%%.

\bibitem{HypExp}
T.~Huber and D.~Maitre, ``{HypExp 2, Expanding Hypergeometric Functions about
  Half-Integer Parameters},''
  \href{http://dx.doi.org/10.1016/j.cpc.2007.12.008}{{\em Comput. Phys.
  Commun.} {\bfseries 178} (2008) 755--776},
\href{http://arxiv.org/abs/0708.2443}{{\ttfamily arXiv:0708.2443 [hep-ph]}}.
%%CITATION = ARXIV:0708.2443;%%.

\bibitem{gradshteyn2007}
I.~S. Gradshteyn and I.~M. Ryzhik, {\em Table of integrals, series, and
  products}.
\newblock Elsevier/Academic Press, Amsterdam, 2007.

\bibitem{Panzer:2014caa}
E.~Panzer, ``{Algorithms for the symbolic integration of hyperlogarithms with
  applications to Feynman integrals},''
  \href{http://dx.doi.org/10.1016/j.cpc.2014.10.019}{{\em Comput. Phys.
  Commun.} {\bfseries 188} (2015) 148--166},
\href{http://arxiv.org/abs/1403.3385}{{\ttfamily arXiv:1403.3385 [hep-th]}}.
%%CITATION = ARXIV:1403.3385;%%.

\bibitem{HPLs}
E.~Remiddi and J.~A.~M. Vermaseren, ``{Harmonic polylogarithms},''
  \href{http://dx.doi.org/10.1142/S0217751X00000367}{{\em Int. J. Mod. Phys.}
  {\bfseries A15} (2000) 725--754},
\href{http://arxiv.org/abs/hep-ph/9905237}{{\ttfamily arXiv:hep-ph/9905237
  [hep-ph]}}.
%%CITATION = HEP-PH/9905237;%%.

\end{thebibliography}\endgroup
\end{document}